\documentclass[11pt]{article}
\usepackage{latexsym,amssymb,amsmath,times}
\usepackage{enumitem}
\usepackage{float}

\usepackage{hyperref}
\usepackage[square,sort&compress,comma,numbers]{natbib}

\makeatletter

\renewcommand{\arraystretch}{2}       
\input amssym.def
\input amssym.tex

\newcommand{\half}{{{\textstyle\frac{1}{2}}}}
\newcommand{\quarter}{{{\textstyle\frac{1}{4}}}}
\newcommand{\be}{\begin{equation}}
\newcommand{\ee}{\end{equation} }
\newcommand{\beqa}{\begin{eqnarray} }
\newcommand{\eeqa}{\end{eqnarray} }
\newcommand{\ba}{\begin{array}}
\newcommand{\ea}{\end{array}}

\newcommand{\Spin}{\mathbf{Spin}}

\newcommand{\Pin}{\mathbf{Pin}}

\newcommand{\deltaS}{\delta_{\varepsilon}}

\newcommand{\ODD}{\mathbf{O}(D,D)}

\newcommand{\Ott}{\mathbf{O}(10,10)}

\newcommand{\Spint}{{\Spin(1,9)}}
\newcommand{\oSpint}{{{\Spin}(9,1)}}
\newcommand{\Pint}{{\Pin(1,9)}}
\newcommand{\oPint}{{{\Pin}(9,1)}}

\newcommand{\brcF}{\bar{\cF}}
\newcommand{\Cp}{{C_{+}}{}}

\newcommand{\brCp}{{\brC_{+}}{}}

\newcommand{\sign}{{\small{\mathbf{{c}}}}}
\newcommand{\signp}{{\small{\mathbf{{c^{\prime}}}}}}

\newcommand{\eleven}{{(11)}}

\newcommand\Tr{{\rm Tr}}

\newcommand\cC{{\cal C}}
\newcommand\cD{{\cal D}}

\newcommand\cF{{\cal F}}

\newcommand\cJ{{\cal J}}

\newcommand\cL{{\cal L}}
\newcommand\cM{{\cal M}}
\newcommand\cN{{\cal N}}

\newcommand\cT{{\cal T}}

\newcommand\tcF{{\tilde{\cal F}}}

\newcommand\hcL{{\hat{\cal L}}}

\newcommand\hGamma{\hat{\Gamma}}

\newcommand\rhop{{\rho^{\prime}}{}}
\newcommand\psip{\psi^{\prime}}
\newcommand\varepsilonp{\varepsilon^{\prime}{}}
\newcommand\brvarepsilon{\bar{\varepsilon}}
\newcommand\brvarepsilonp{\brvarepsilon^{\prime}{}}
\newcommand\brrhop{\brrho^{\prime}{}}
\newcommand\brpsip{\brpsi^{\prime}}

\newcommand\aap{a^{\prime}}
\newcommand\bbp{b^{\prime}}
\newcommand\ccp{c^{\prime}}
\newcommand\ddp{d^{\prime}}

\newcommand\typeT{{\scriptscriptstyle{\rm Type\,II}}}

\newcommand\dis{\displaystyle}

\newcommand\seceq{\simeq}

\def\bra{\bar{a}}
\def\brb{\bar{b}}
\def\brc{\bar{c}}
\def\bre{\bar{e}}

\def\breta{\bar{\eta}}
\def\bralpha{\bar{\alpha}}
\def\brbeta{\bar{\beta}}
\def\brgamma{\bar{\gamma}}
\def\brdelta{\bar{\delta}}
\def\brrho{\bar{\rho}}
\def\brpsi{\bar{\psi}}

\def\brp{{\bar{p}}}
\def\brq{{\bar{q}}}
\def\brr{{\bar{r}}}
\def\brs{{\bar{s}}}
\def\brt{{\bar{t}}}

\def\brPhi{{{\bar{\Phi}}}}
\def\brDelta{{{\bar{\Delta}}}}

\def\brC{\bar{C}}
\def\brF{\bar{F}}

\def\brV{{\bar{V}}}
\def\brP{{\bar{P}}}

\def\Tw{{T}}

\newcommand{\na}{{\nabla}}

\def\Gammao{\Gamma^{\scriptscriptstyle{0}}}
\def\Phio{\Phi^{\scriptscriptstyle{0}}}

\def\brPhio{\brPhi^{\scriptscriptstyle{0}}}

\def\cDo{\cD^{\scriptscriptstyle{0}}}

\def\So{S^{\scriptscriptstyle{0}}}

\def\hcD{\hat{\cD}}

\def\hcDp{\hat{\cD}{}^{\prime}}

\def\tcD{\tilde{\cD}}
\def\tGamma{\tilde{\Gamma}}
\def\tcDp{\tilde{\cD}{}^{\prime}}
\def\tGammap{\tilde{\Gamma}{}^{\prime}}

\newcommand\cDsh{\cD^{\sharp}}
\newcommand\cDfl{\cD^{\flat}}
\newcommand\Gammash{\Gamma^{\sharp}}
\newcommand\Gammafl{\Gamma^{\flat}}

\newcommand\cDpsh{\cD^{\prime\sharp}}

\newcommand\cDpfl{\cD^{\prime\flat}}

\newcommand\Gammapfl{\Gamma^{\prime\flat}}
\newcommand\Gammapsh{\Gamma^{\prime\sharp}}

\newcommand\cDs{\cD^{\star}}
\newcommand\cDps{\cD^{\prime\star}}

\def\Gammas{\Gamma^{\star}}
\def\Gammaps{\Gamma^{\prime\star}}

\begin{document}
\begin{titlepage}
\title{
Stringy  Unification of  Type IIA and IIB Supergravities under\\ ${\cN=2}$ ${D=10}$   Supersymmetric  Double Field Theory\\~\\}
\author{\sc Imtak Jeon,${}^{\sharp}$\mbox{~}  
Kanghoon Lee,${}^{\flat}$\mbox{~} 
Jeong-Hyuck Park${}^{\dagger}$\mbox{~}   
and\mbox{~}   Yoonji Suh${}^{\dagger}$ }
\date{}
\maketitle \vspace{-1.0cm}
\begin{center}
~~~\\
${}^{\sharp}$Harish-Chandra Research Institute, Chhatnag Road, Jhusi, Allahabad 211019, India\\
\texttt{imtak110@gmail.com}\\
~\\
${}^{\flat}$Center for Quantum Spacetime, Sogang University,  Seoul 121-742, Korea\\
\texttt{kanghoon@sogang.ac.kr}\\
~\\
${}^{\dagger}$Department of Physics, Sogang University,  Seoul 121-742, Korea\\
\texttt{park@sogang.ac.kr,\quad yjsuh@sogang.ac.kr}
~{}\\
~~~\\~\\
\end{center}
\begin{abstract}
\vskip0.2cm
\noindent To the full order in fermions, we construct ${D=10}$ type II   supersymmetric double field theory. We  spell   the  precise   ${\cN=2}$ supersymmetry transformation rules as for $32$ supercharges.      The constructed action unifies type IIA and IIB supergravities in a  manifestly covariant manner  with respect to     $\Ott$ T-duality and    a pair of  local Lorentz groups, or  $\Spint\times\oSpint$,  besides   the usual general covariance  of  supergravities   or the generalized diffeomorphism. While the theory is unique, the solutions are twofold. Type  IIA and IIB supergravities are   identified as two different types of solutions rather than two different theories.
\end{abstract}

{\small
\begin{flushleft}
~~
\end{flushleft}}
\thispagestyle{empty}
\end{titlepage}
\newpage
\section*{Introduction} 
Strings perceive spacetime   in a  different   way than  particles do through Riemannian geometry.  While the fundamental object in Riemannian geometry is the metric,  string theory  puts the  Kalb-Ramond $B$-field and a scalar dilaton on an equal footing  along  with the metric, since they form a multiplet of T-duality~\cite{Buscher:1985kb,Buscher:1987sk,Buscher:1987qj},  a genuine  stringy  property  which is not present  in ordinary particle  theory.  \\  

\noindent Although type IIA and IIB supergravities  provide   low energy effective descriptions of closed superstrings, once formulated  within  the  Riemannian setup,  they appear unable    to  capture the full stringy structure like  T-duality  or to explain the appearance of enhanced symmetries  after dimensional reductions~\cite{Giveon:1988tt,Meissner:1991zj}. String theory seems to  urge      us to look for  a novel mathematical framework, such as Generalized Geometry~\cite{Gualtieri:2003dx,Hitchin:2004ut,Hitchin:2010qz}  or Double Field Theory (DFT)~\cite{Hull:2009mi,Hull:2009zb,Hohm:2010jy,Hohm:2010pp} (see also \cite{Siegel:1993xq,Siegel:1993th} for   relevant pioneering     works). \\

\noindent While generalized geometry   combines   tangent and cotangent spaces    giving    a  geometric meaning to the $B$-field~\cite{Grana:2008yw,Koerber:2010bx}, DFT      doubles    the spacetime dimension,  from $D$ to ${D+D}$ in order to manifest the $\ODD$ T-duality group structure~\cite{Tseytlin:1990nb,Tseytlin:1990va,Siegel:1993xq,Siegel:1993th}.   With an additional requirement of   so called  \textit{strong constraint} or \textit{section condition},  DFT reduces to a  known   string theory effective action in $D$-dimension.  The section condition means     that all the DFT-fields   live on a $D$-dimensional null hyperplane such that, the  $\ODD$ invariant d'Alembertian operator is  trivial acting on arbitrary fields as well as their products,
\be
\ba{ll}
\partial_{A}\partial^{A}=\cJ^{AB}\partial_{A}\partial_{B}\seceq 0\,,~~&~~\cJ^{AB}=\left(\ba{cc}0&1\\1&0\ea\right)\,.
\ea
\label{constraint}
\ee
~\\
\noindent  DFT unifies  the $B$-field  gauge symmetry and   the  diffeomorphism, as  both  are generated by   generalized Lie derivative~\cite{Courant,Gualtieri:2003dx} (see also \cite{Hohm:2012gk} for finite transformations),
\be
\hcL_{X}\Tw_{A_{1}\cdots A_{n}}:=X^{B}\partial_{B}\Tw_{A_{1}\cdots A_{n}}+\omega_{{\scriptscriptstyle{T\,}}}\partial_{B}X^{B}\Tw_{A_{1}\cdots A_{n}}
+\sum_{i=1}^{n}(\partial_{A_{i}}X_{B}-\partial_{B}X_{A_{i}})\Tw_{A_{1}\cdots A_{i-1}}{}^{B}{}_{A_{i+1}\cdots  A_{n}}\,.
\label{tcL}
\ee
~\\
\noindent  Further,  recent study   of  the  Scherk-Schwarz reduction in  DFT  has  shown  that,   by relaxing  the section condition~(\ref{constraint})  ---and hence in a truly non-Riemannian set up---   one may  derive  all the known gauged supergravities in lower than ten dimensions~\cite{Geissbuhler:2011mx,Aldazabal:2011nj,Grana:2012rr,Dibitetto:2012rk,Dibitetto:2012xd}.   This seems to  indicate the potential power of DFT and motivates further explorations.    \\ \newpage

\noindent In this work,  we construct ${\cN=2}$  ${D=10}$ supersymmetric double field theory (SDFT). We carry out the construction  employing    
\textit{genuine  SDFT  field-variables} which are subject to the section condition~(\ref{constraint}) and    differ \textit{a priori}  from Riemannian, or supergravity     variables. For example,  ordinary zehnbeins   and various  form-fields will  never  enter  in our  construction.   We tend to believe that   the  usage of the  genuine SDFT field-variables is quite crucial  and it  essentially  ensures    the   following  properties of the final results.

\begin{itemize}
\item Each term in the  constructed Lagrangian  is  manifestly and simultaneously  covariant with respect to  $\Ott$ T-duality,   a pair of  local Lorentz groups,   ${\Spint\times\oSpint}$,  and   the  DFT-diffeomorphism generated by  $\hcL_{X}$ in (\ref{tcL}).   
\item The supersymmetric completion is fulfilled  to the full order in fermions.   
\item Further,  ${\cN=2}$  ${D=10}$ SDFT unifies type IIA and IIB supergravities: while the theory is unique,  the solutions are twofold,   type  IIA and type  IIB.  
\end{itemize}

Related key precedents include Refs.\cite{Coimbra:2011nw,Coimbra:2012yy,Hohm:2011zr,Hohm:2011dv}. In  \cite{Coimbra:2011nw,Coimbra:2012yy}, within the  generalized geometry setup in terms of a pair of zehnbeins and various  form-fields, 
the type II supergravity was reformulated  into an $\Spint\times\oSpint$ covariant form (up to quadratic order in fermions). In  \cite{Hohm:2011zr,Hohm:2011dv},  the bosonic part of type II  SDFT was proposed which in particular put  the \mbox{R-R} sector in an $\Ott$ spinorial representation, as in \cite{Fukuma:1999jt,Hassan:1999mm}.    In our case, the  R-R sector is in a  ${\Spint\times\oSpint}$ bi-fundamental spinorial representation, \textit{e.g.~}`$\,\cC^{\alpha}{}_{\bralpha}\,$'. Table~\ref{TABindices} summarizes our  index gymnastics. 
  
{\scriptsize{\begin{table}[H]
\begin{center}
\begin{tabular}{c|c|c}
Index~&~Representation~&Raising \& Lowering Indices\\
\hline
$A,B,\cdots$&$\Ott$ \& $\hcL_{X}$ vector&$\cJ_{AB}$\\
$p,q,\cdots$&$\Pint$  vector~&$\eta_{pq}=\mbox{diag}(-++\cdots+)$ \\
$\alpha,\beta,\cdots$&$\Pint$  spinor~&$\Cp_{\alpha\beta}$\,,~~$(\gamma^{p})^{T}=C_{+}\gamma^{p}C_{+}^{-1}$\\
$\brp,\brq,\cdots$&$\oPint$  vector~&$\breta_{\brp\brq}=\mbox{diag}(+--\cdots-)$ \\
$\bralpha,\brbeta,\cdots$&$\oPint$  spinor~&$\brCp_{\bralpha\brbeta}$\,,~~$(\brgamma^{\brp})^{T}=\brC_{+}\brgamma^{\brp}\brC_{+}^{-1}$\\
\end{tabular}
\caption{Index  for  each symmetry representation and the corresponding ``metric" to raise or lower the positions.  For further details and a review on the  formalism,  we refer the reader to the Appendix of \cite{Jeon:2012kd}. } 
\label{TABindices}
\end{center}
\end{table}}}
\section*{Field Content} 
We postulate \textit{the fundamental fields of type II SDFT} to be strictly,  from \cite{Jeon:2010rw,Jeon:2011kp,Jeon:2011cn,Jeon:2011vx,Jeon:2011sq,Jeon:2012kd},  
\be
\ba{llllllll}
d\,,&~V_{Ap}\,,&~\brV_{A\brp}\,,&~\cC^{\alpha}{}_{\bralpha}\,,&~
\psi^{\alpha}_{\brp}\,,&~\rho^{\alpha}\,,&~\psi^{\prime\bralpha}_{p}\,,&~\rho^{\prime\bralpha}\,.
\ea
\label{FFC}
\ee
We wish to  stress that,   for   the sake of the   full covariance and the (relatively)  compact way of  full order  supersymmetric completion,  it is crucial   to set the fundamental fields to be precisely those above. Although some of them may be parametrized in terms of Riemannian zehnbeins and form-fields, the parametrization is not unique, may render ``non-geometric" interpretations,   and will certainly   becloud  the whole   symmetry structure listed  in Table~\ref{TABindices}.  \\

\noindent  Firstly  for the NS-NS sector, the DFT-dilaton, $d$,  gives rise to a scalar density  with  weight one, $e^{-2d}$~\cite{Hull:2009zb}.  The DFT-vielbeins, $V_{Ap}$, $\brV_{A\brp}$,  satisfy the following  four   \textit{defining properties}~\cite{Jeon:2011cn,Jeon:2011vx}:
\be
\ba{ll}
V_{Ap}V^{A}{}_{q}=\eta_{pq}\,,\quad&\quad
\brV_{A\brp}\brV^{A}{}_{\brq}=\breta_{\brp\brq}\,,\\
V_{Ap}\brV^{A}{}_{\brq}=0\,,\quad&\quad V_{Ap}V_{B}{}^{p}+\brV_{A\brp}\brV_{B}{}^{\brp}=\cJ_{AB}\,.
\ea
\label{defV}
\ee
In particular, they  generate a pair of orthogonal and complete projections, 
\be
\ba{ll}
P_{AB}=V_{A}{}^{p}V_{Bp}\,,\quad&\quad
\brP_{AB}=\brV_{A}{}^{\brp}\brV_{B\brp}\,,
\ea
\ee
satisfying 
\be
\ba{llll}
P_{A}{}^{B}P_{B}{}^{C}=P_{A}{}^{C}\,,~~~~&~~~~\brP_{A}{}^{B}\brP_{B}{}^{C}=\brP_{A}{}^{C}\,,\quad&\quad
P_{A}{}^{B}\brP_{B}{}^{C}=0\,,~~~~&~~~~P_{A}{}^{B}+\brP_{A}{}^{B}=\delta_{A}{}^{B}\,.
\ea
\label{projection}
\ee
The DFT-vielbeins, $V_{Ap}$, $\brV_{A\brp}$, are $\ODD$ vectors as the  index structure indicates.  They are   the only    field variables  in (\ref{FFC}) which are  $\ODD$ non-singlet.  As a solution to  (\ref{defV}),  they can be  parametrized     in terms of ordinary    zehnbeins  and $B$-field,   in various ways up to $\ODD$ rotations and field redefinitions~\cite{Jeon:2012kd}. Yet,  in  order to maintain the clear  manifestation    of the  $\ODD$ covariance,  it is necessary  to work with   the parametrization-independent and  $\ODD$ covariant  DFT-vielbeins, \textit{i.e.~}$V_{Ap}$ and $\brV_{A\brp}$,  rather than the Riemannian variables, \textit{i.e.~}ordinary zehnbeins  and $B$-field. \\

\noindent  For fermions,   the gravitinos  and  the DFT-dilatinos  are    \textit{not twenty, but ten-dimensional} Majorana-Weyl spinors, as in \cite{Coimbra:2011nw,Coimbra:2012yy},
\be
\ba{ll}
{\gamma^{\eleven}\psi_{\brp}}=\sign\,\psi_{\brp}\,,\quad &\quad
\gamma^{\eleven}\rho=-\sign\,\rho\,,\\
\brgamma^{\eleven}\psi^{\prime}_{p}=\signp\psi^{\prime}_{p}\,,\quad&\quad
\brgamma^{\eleven}\rhop=-\signp\rhop\,,
\ea
\label{chiralityF}
\ee
where $\sign$ and $\signp$ are arbitrary  independent two sign factors, $\sign^{2}=\signp{}^{2}=1$. Yet, \textit{a priori}   all the possible four  different sign choices are   {equivalent} up to   $\Pint\times\oPint$ rotations. That is to say,  $\cN=2$ $D=10$ SDFT is  \textit{chiral} with respect to  \textit{both} $\Pint$ {and} $\oPint$, and  the theory is unique.  Hence, without loss of generality, we may safely set
${\sign\equiv\signp\equiv+1}$.  Later we shall see    that, while  the theory is unique  the solutions are twofold and can be identified as type IIA or IIB supergravity backgrounds.   \\
\indent We also have ${\cN=2}$ supersymmetry parameters, $\varepsilon$, $\varepsilonp$,  which carry  the same chirality as the gravitinos,  such that $ \gamma^{\eleven}\varepsilon=\sign\,\varepsilon$, $\brgamma^{\eleven}\varepsilonp=\signp\varepsilonp$.\\

\noindent  Lastly for the R-R sector,  we set  the R-R potential, $\cC^{\alpha}{}_{\bralpha}$,  to be  in the  bi-fundamental  spinorial   representation of  $\Pint\times\oPint$~\cite{Coimbra:2011nw,Coimbra:2012yy,Jeon:2012kd} rather than an $\Ott$ spinorial one~\cite{Hohm:2011zr,Hohm:2011dv}.  It  possesses   the    chirality,
\be
\gamma^{\eleven}\cC\brgamma^{\eleven}=\sign\signp\,\cC\,.
\label{RRD}
\ee


\section*{Derivatives}  
Another essential ingredient  is so called   \textit{master semi-covariant  derivative} from \cite{Jeon:2011vx},
\be
\cD_{A}=\partial_{A}+\Gamma_{A}+\Phi_{A}+\brPhi_{A}\,,
\label{MSCD}
\ee 
which  contains generically  three kinds of  connections:  $\Gamma_{A}$ for  the   DFT-diffeomorphism or the generalized Lie derivative~(\ref{tcL}),  $\Phi_{A}$ for   $\Spint$ and  $\brPhi_{A}$  for  $\oSpint$  local Lorentz symmetries.  Contracted with the projections~(\ref{projection})  or the DFT-vielbeins properly,   it can produce various  fully covariant derivatives, and hence the name, `semi-covariant'~\cite{Jeon:2011cn,Jeon:2011vx,Jeon:2012kd}.

By definition, the master derivative~(\ref{MSCD}) is   required to be   compatible with all the constants  in Table~\ref{TABindices} (``metrics" and gamma matrices), and further   to annihilate the whole NS-NS sector, 
\be
\ba{lll}
\cD_{A}d=0\,,~~&~~\cD_{A}V_{Bp}=0\,,~~&~~\cD_{A}\brV_{A\brp}=0\,.
\ea
\ee
The connections are then related to each other through
\be
\ba{lll}
\Phi_{Apq}=V^{B}{}_{p}\na_{A}V_{Bq}\,,\quad&\quad
\brPhi_{A\brp\brq}=\brV^{B}{}_{\brp}\na_{A}\brV_{B\brq}\,,\quad&\quad
\Gamma_{ABC}=V_{B}{}^{p}D_{A}V_{Cp}+\brV_{B}{}^{\brp}D_{A}\brV_{C\brp}\,,
\ea
\label{c3}
\ee
where we put $\na_{A}=\partial_{A}+\Gamma_{A}$ and $D_{A}=\partial_{A}+\Phi_{A}+\brPhi_{A}$.  \\
\indent Especially,   as the DFT analogy of the Riemannian    Christoffel connection, the \textit{torsionless connection}, $\Gammao_{A}$, can be uniquely singled out~\cite{Jeon:2011cn,Jeon:2012kd} (\textit{c.f.~}\cite{Hohm:2011si}):
\be
\ba{ll}
\Gammao_{CAB}=&2(P\partial_{C}P\brP)_{[AB]}
+2({{\brP}_{[A}^{~D}{\brP}_{B]}^{~E}}-{P_{[A}^{~D}P_{B]}^{~E}})\partial_{D}P_{EC}\\
{}&-\textstyle{\frac{4}{9}}(\brP_{C[A}\brP_{B]}^{~D}+P_{C[A}P_{B]}^{~D})(\partial_{D}d+(P\partial^{E}P\brP)_{[ED]}),
\ea
\label{Gammao}
\ee
such that a generic torsionful  DFT-diffeomorphism  connection  assumes the following  general form:
\be
\Gamma_{CAB}=\Gammao_{CAB}+\Delta_{C[pq]}V_{A}{}^{p}V_{B}{}^{q}+\brDelta_{C[\brp\brq]}\brV_{A}{}^{\brp}\brV_{B}{}^{\brq}\,,
\label{PhibrPhi}
\ee
where $\Delta_{C[pq]}$ and  $\brDelta_{C[\brp\brq]}$ correspond to torsions. Explicitly we shall employ  four different kinds of torsions:  (\ref{Gamma1.5}) for the curvature,  (\ref{GsGps}) for the fermionic kinetic terms,   (\ref{Ghat}) for the supersymmetry, and (\ref{GGG}) for the equations of motion.\\

\indent The \textit{R-R field strength}, $\cF^{\alpha}{}_{\bralpha}$,  is defined from \cite{Jeon:2012kd},
\be
\cF:=\cDo_{+}\cC\,,
\label{RRFLUX}
\ee
where  $\cDo_{+}$ corresponds to one of the two fully covariant and \textit{nilpotent}  differential operators, $\cDo_{\pm}$, which are  set by the torsionless connection~(\ref{Gammao}), and  may  act  on an arbitrary 
$\Pint\times\oPint$ bi-fundamental field, $\cT^{\alpha}{}_{\brbeta}$\,: 
\be
\ba{ll}
\cDo_{\pm}\cT:=\gamma^{p}\cDo_{p}\cT\pm\gamma^{\eleven}\cDo_{\brp}\cT\brgamma^{\brp}\,,~&~
(\cDo_{\pm})^{2}\cT\seceq0\,,
\ea
\label{Dpm}
\ee
where we put\footnote{Strictly speaking,   due to the presence of  $\gamma^{\eleven}$  in (\ref{Dpm}), the R-R field strength,  $\cF=\cDo_{+}\cC$,  is  covariant ---up to the flipping of the chirality--- with respect to, not  $\Pint\times\oPint$ but $\Spint\times\oPint$.   For the opposite equivalent choice,  see  eq.(2.25) in \cite{Jeon:2012kd}.}
 $\cDo_{p}=V^{A}{}_{p}\cDo_{A}$ and $\cDo_{\brp}=\brV^{A}{}_{\brp}\cDo_{A}$.


\section*{Curvature}  
The final ingredient we shall employ     is   the \textit{semi-covariant DFT-curvature,}  $S_{ABCD}$, from \cite{Jeon:2011cn},  
\be
S_{ABCD}:=\half\left(R_{ABCD}+R_{CDAB}-\Gamma^{E}{}_{AB}\Gamma_{ECD}\right)\,,
\label{SABCD}
\ee  
which is defined through the standard  (yet never-covariant) field strength of the  DFT-diffeomorphism connection~(\ref{PhibrPhi}), 
\be
R_{CDAB}=\partial_{A}\Gamma_{BCD}-\partial_{B}\Gamma_{ACD}+\Gamma_{AC}{}^{E}\Gamma_{BED}-\Gamma_{BC}{}^{E}\Gamma_{AED}\,.
\label{RABCD}
\ee
Again, with the help of the   projections,  it  can produce  fully covariant  curvatures, such as    Ricci~(\ref{RicciS}) and  scalar,    
\be
(P^{AB}P^{CD}-\brP^{AB}\brP^{CD})S_{ACBD}\,.
\ee


\section*{The Lagrangian  and  Supersymmetry} 
The Lagrangian   of ${\cN=2}$ ${D=10}$   SDFT we  construct  in this work   is the following,
\be
\ba{l}
\cL_{\typeT}=e^{-2d}\Big[\textstyle{\frac{1}{8}}(P^{AB}P^{CD}-\brP^{AB}\brP^{CD})S_{ACBD}+\half\Tr(\cF\bar{\cF})-i\brrho\cF\rhop+i\brpsi_{\brp}\gamma_{q}\cF\brgamma^{\brp}\psip{}^{q}\\
\,\quad\quad\quad\quad\quad\quad\quad+i\half\brrho \gamma^{p}\cDs_{p}\rho
-i\brpsi^{\brp}\cDs_{\brp}\rho
-i\half\brpsi^{\brp}\gamma^{q}\cDs_{q}\psi_{\brp}
-i\half\brrhop \brgamma^{\brp}\cDps_{\brp}\rhop
+i\brpsip{}^{p}\cDps_{p}\rhop+i\half\brpsip{}^{p}\brgamma^{\brq}\cDps_{\brq}\psip_{p}
\Big]\,.
\ea
\label{typeIIL}
\ee
As they are contracted with  the DFT-vielbeins properly,  \textit{each   term  in the Lagrangian  is fully covariant} with respect to $\Ott$  T-duality, ${\Spint\times\oSpint}$ local Lorentz symmetry and   the  DFT-diffeomorphism. With  the charge conjugation of the R-R field strength, $\bar{\cF}=\brC_{+}^{-1}\cF^{T}C_{+}$, the trace, $\Tr(\cF\bar{\cF})$ in  (\ref{typeIIL}) is over the $\Spint$ spinorial indices. \\
~\\

\noindent  The $\cN=2$  \textit{supersymmetry transformation rules} are
\be
\ba{l}
\deltaS  d=-i\half(\brvarepsilon\rho+\brvarepsilonp\rhop)\,,\\
\deltaS V_{Ap}=i\brV_{A}{}^{\brq}(\brvarepsilonp\bar{\gamma}_{\brq}\psi^{\prime}_{p}
-\bar{\varepsilon}\gamma_{p}\psi_{\brq})\,,\\
\deltaS \brV_{A\brp}=iV_{A}{}^{q}(
\bar{\varepsilon}\gamma_{q}\psi_{\brp}-\brvarepsilonp\bar{\gamma}_{\brp}\psi^{\prime}_{q})\,,\\
\deltaS  \cC= i \half 
(\gamma^{p}\varepsilon \brpsi^{\prime}_{p}-\varepsilon\brrhop-\psi_{\brp}\brvarepsilonp\bar{\gamma}^{\brp}+\rho\brvarepsilonp)+\cC\deltaS d 
-\half(\bar{V}^{A}{}_{\brq\,}\deltaS  V_{Ap})\gamma^{(d+1)}\gamma^{p}\cC \bar{\gamma}^{\brq}\,,\\
\deltaS  \rho= -\gamma^{p}\hcD_{p}\varepsilon + i\half \gamma^{p}\varepsilon\, \brpsi^{\prime}_{p}\rhop -i \gamma^{p} \psi^{\brq}\brvarepsilonp \brgamma_{\brq} \psi^{\prime}_{p}\,,
\\
\deltaS  \rhop= -\bar{\gamma}^{\brp}\hcDp_{\brp}\varepsilonp + i \half \brgamma^{\brp}\varepsilonp \, \brpsi_{\brp} \rho -i \brgamma^{\brq}\psi^{\prime}_{p} \brvarepsilon \gamma^{p} \psi_{\brq}\,,
\\
\deltaS  \psi_{\brp}=\hcD_{\brp}\varepsilon+(\cF -i\frac{1}{2} \gamma^{q}\rho\,\brpsi^{\prime}_{q} +i\half \psi^{\brq}\, \brrhop\brgamma_{\brq} ) \brgamma_{\brp} \varepsilonp +i\frac{1}{4} \varepsilon \brpsi_{\brp} \rho + i \half \psi_{\brp} \brvarepsilon \rho\,,
\\
\deltaS  \psi^{\prime}_{p}=\hcD^{\prime}_{p}\varepsilonp +( \bar{\cF} - i \frac{1}{2} \brgamma^{\brq}\rhop\brpsi_{\brq} +i \half \psi^{\prime q}  \brrho \gamma_{q}) \gamma_{p}\varepsilon +i\frac{1}{4} \varepsilonp \brpsi^{\prime}_{p} \rhop + i \half \psi^{\prime}_{p} \brvarepsilon^{\prime}\rhop\,.
\ea
\label{N2SUSY}
\ee
\newpage

\section*{Torsions}  
Presenting our main results above, (\ref{typeIIL}) and (\ref{N2SUSY}),  we have organized  all the higher order fermionic terms into various torsions.   Firstly,  with (\ref{SABCD}), the DFT-curvature, $S_{ABCD}$, in  the Lagrangian  is given   by the connection,
\be
\ba{ll}
\Gamma_{ABC}=&\Gammao_{ABC}+i\textstyle{\frac{1}{3}}\brrho\gamma_{ABC}\rho-2i\brrho\gamma_{BC}\psi_{A}
-i\textstyle{\frac{1}{3}}\brpsi^{\brp}\gamma_{ABC}\psi_{\brp}+4i\brpsi_{B}\gamma_{A}\psi_{C}\\
{}&\quad+i\textstyle{\frac{1}{3}}\brrhop\brgamma_{ABC}\rhop
-2i\brrhop\brgamma_{BC}\psip_{A}-i\textstyle{\frac{1}{3}}\brpsi^{\prime p}\brgamma_{ABC}\psip_{p}+4i\brpsip_{B}\brgamma_{A}\psip_{C}\,.
\label{Gamma1.5}
\ea
\ee
Secondly, the master derivatives in the  fermionic kinetic terms   are twofold:   $\cDs_{A}$ for  the unprimed fermions and $\cDps_{A}$ for the  primed fermions. They are   set by the following twin connections,
\be
\ba{ll}
\Gammas_{ABC}=&\Gamma_{ABC}-i\textstyle{\frac{11}{96}}\brrho\gamma_{ABC}\rho+i\textstyle{\frac{5}{4}}\brrho\gamma_{BC}\psi_{A}+i\textstyle{\frac{5}{24}}\brpsi^{\brp}\gamma_{ABC}\psi_{\brp}-2i\brpsi_{B}\gamma_{A}\psi_{C}+i\textstyle{\frac{5}{2}}\brrhop\brgamma_{BC}\psip_{A}\,,\\
\Gammaps_{ABC}=&\Gamma_{ABC}-i\textstyle{\frac{11}{96}}\brrhop\brgamma_{ABC}\rhop+i\textstyle{\frac{5}{4}}\brrhop\brgamma_{BC}\psip_{A}+i\textstyle{\frac{5}{24}}\brpsi^{\prime p}\brgamma_{ABC}\psip_{p}-2i\brpsip_{B}\brgamma_{A}\psip_{C}+i\textstyle{\frac{5}{2}}\brrho\gamma_{BC}\psi_{A}\,.
\ea
\label{GsGps}
\ee
Similarly, for  the supersymmetry transformations~(\ref{N2SUSY}),   we take
\be
\ba{ll}
\hGamma_{ABC}=&\Gamma_{ABC}
-i\textstyle{\frac{17}{48}}\brrho\gamma_{ABC}\rho+i\textstyle{\frac{5}{2}}\brrho\gamma_{BC}\psi_{A}+i\textstyle{\frac{1}{4}}\brpsi^{\brp}\gamma_{ABC}\psi_{\brp} -3i \brpsi^{\prime}_{B}
\brgamma_{A} \psi^{\prime}_{C}\,,\\
\hGamma^{\prime}_{ABC}=&\Gamma_{ABC}
-i\textstyle{\frac{17}{48}}\brrhop\brgamma_{ABC}\rhop+i\textstyle{\frac{5}{2}}\brrhop\brgamma_{BC}\psip_{A}+i\textstyle{\frac{1}{4}}\brpsi^{\prime p}\brgamma_{ABC}\psip_{p}-3i \brpsi_{B}\gamma_{A} \psi_{C}\,.
\ea
\label{Ghat}
\ee
The  connection, $\Gamma_{ABC}$ given in (\ref{Gamma1.5}) and also appearing in  (\ref{GsGps}),  (\ref{Ghat}),  has been  fixed  by requiring the 1.5 formalism to work,  see (\ref{varL}). The additional  parts of the  connections in   (\ref{GsGps}) and  (\ref{Ghat})  are then uniquely determined  from the full order supersymmetric completion.\\ 

\section*{Self-duality and Equations of Motion} 
{The type II SDFT Lagrangian    (\ref{typeIIL}) is {pseudo}:} An additional \textit{self-duality} relation needs to be imposed by hand  on the R-R field strength combined  with fermions,
\be
\tcF_{-}:=\big(1-\gamma^{\eleven}\big)\left(\cF-i\half\rho\brrhop+i\half\gamma^{p}\psi_{\brq}\brpsi^{\prime}_{p}\brgamma^{\brq}\right)\equiv 0\,.
\label{SD}
\ee
~\\
\noindent Under arbitrary  infinitesimal  variations of all the fields, the Lagrangian transforms, up to total derivatives,  
\be
\ba{ll}
\delta\cL_{\typeT}\seceq&-2\delta d\times\cL_{\typeT}\\
{}&+\delta\Gamma_{ABC}\times 0\\
{}&+\half e^{-2d} \delta V^{Bp}\brV_{B}{}^{\brq}\left[\tilde{S}_{p\brq}
+\Tr(\cF\brgamma_{\brq}\brcF\gamma_{p})
\right]\\
{}&-ie^{-2d}\widetilde{\delta\brpsi^{\brp}}
\left(\tcD_{\brp}\rho+\gamma^{p}\tcD_{p}\psi_{\brp}-\gamma^{p}\cF\brgamma_{\brp}\psip_{p}\right)\\
{}&+ie^{-2d}\widetilde{\delta\brrho\,}\left(\gamma^{p}\tcD_{p}\rho-\tcD_{\brp}\psi^{\brp}-\cF\rho^{\prime}\right)\\
{}&+ie^{-2d}\widetilde{\delta\brpsi^{\prime p}}
\left(\tcDp_{p}\rhop+\brgamma^{\brp}\tcDp_{\brp}\psip_{p}-\brgamma^{\brp}\brcF\gamma_{p}\psi_{\brp}\right)\\
{}&-ie^{-2d}\widetilde{\delta\brrhop}
\left(\brgamma^{\brp}\tcDp_{\brp}\rhop-\tcDp_{p}\psi^{\prime p}-\brcF\rho\right)\\
{}&+e^{-2d}\Tr\!\left[\tcF_{-}\left(\delta d\brcF-\half 
\delta V^{Ap}\brV_{A}{}^{\brq}\brgamma_{\brq}\brcF\gamma_{p}\right)-\cDo_{-}\tcF_{-}\overline{\widetilde{\delta\cC}}\, \right]\,.
\ea
\label{varL}
\ee
Each line then corresponds to  the  equation of motion  of ${\cN=2}$ ${D=10}$ SDFT.  
 In particular, the on-shell Lagrangian vanishes, ${\cL_{\typeT}=0}$, and the  DFT-generalization of the \textit{Einstein equation} follows
\be
\tilde{S}_{p\brq}+\Tr(\cF\brgamma_{\brq}\brcF\gamma_{p})=0\,.
\ee
The self-duality~(\ref{SD}) implies the equation of motion for the  R-R potential,  $\cDo_{-}\tcF_{-}=0$. Further,  as in the \mbox{$\cN=1$} SDFT~\cite{Jeon:2011sq}, the 1.5 formalism, `$\delta\Gamma_{ABC}\times 0$',  nicely works here  with   the connection spelled  in (\ref{Gamma1.5}). \\

\noindent Writing  (\ref{varL}),  we set  some shorthand notations: For the arbitrary variations of the fields,
\be
\ba{l}
\widetilde{\delta\brrho\,}:=\delta\brrho-\textstyle{\frac{1}{4}}\delta V_{Bq}\brrho\gamma^{Bq}\,,\\
\widetilde{\delta\brpsi^{\brp}}:=\delta\brpsi^{\brp}-\delta\brV^{B\brp} \brpsi_{B}
-\textstyle{\frac{1}{4}}\delta V_{Bq}\brpsi^{\brp}\gamma^{Bq}\,,\\
\widetilde{\delta\brrhop}:=\delta\brrhop-\textstyle{\frac{1}{4}}\delta \brV_{B\brq}\brrhop\brgamma^{B\brq}\,,\\
\widetilde{\delta\brpsi^{\prime p}}:=
\delta\brpsi^{\prime p}-\delta V^{Bp}\brpsip_{B}-\!\textstyle{\frac{1}{4}}\delta \brV_{B\brq}\brpsi^{\prime p}\brgamma^{B\brq}\,,\\
\widetilde{\delta\cC}:=\delta\cC -\cC\delta d +\quarter \delta V_{Ap}\gamma^{Ap}\cC-\quarter\delta\brV_{A\brp}\cC\brgamma^{A\brp}+\half\delta V_{Ap}\gamma^{\eleven}\gamma^{p}\cC\brgamma^{A}\,,
\ea
\ee
and  for the    Ricci  curvature,
\be
\tilde{S}_{p\brq}:=V^{A}{}_{p}\brV^{B}{}_{\brq}S^{C}{}_{ACB}+2i\brpsi_{\brq}\tcD_{p}\rho-i\brpsi^{\brp}\gamma_{p}\tcD_{\brq}\psi_{\brp}+2i\brpsip_{p}\tcDp_{\brq}\rhop-i\brpsi^{\prime q}\brgamma_{\brq}\tcDp_{p}\psip_{q}+i\brrho\gamma_{p}\tcD_{\brq}\rho+i\brrhop\brgamma_{\brq}\tcDp_{p}\rhop\,.
\label{RicciS}
\ee
We also  set   the derivatives, $\tcD_{A}, \tcD_{A}^{\prime}$ appearing \mbox{in (\ref{varL}),} by
\be
\ba{ll}
\tGamma_{ABC}=&\Gamma_{ABC}
-i\textstyle{\frac{23}{54}}\brrho\gamma_{ABC}\rho
+i\textstyle{\frac{23}{27}}\brrho\gamma_{BC}\psi_{A}
+i\textstyle{\frac{23}{54}}\brpsi^{\brp}\gamma_{ABC}\psi_{\brp}
-i\textstyle{\frac{73}{18}}\brpsi_{B}\gamma_{A}\psi_{C}\\
{}&-i\textstyle{\frac{5}{4}}\brrhop\brgamma_{ABC}\rhop
+i\textstyle{\frac{5}{2}}\brrhop\brgamma_{BC}\psip_{A}
+i\textstyle{\frac{5}{4}}\brpsi^{\prime p}\brgamma_{ABC}\psip_{p}
-5i\brpsip_{B}\brgamma_{A}\psip_{C}\,,\\
\tGammap_{ABC}=&\Gamma_{ABC}
-i\textstyle{\frac{23}{54}}\brrhop\brgamma_{ABC}\rhop
+i\textstyle{\frac{23}{27}}\brrhop\brgamma_{BC}\psip_{A}
+i\textstyle{\frac{23}{54}}\brpsi^{\prime p}\brgamma_{ABC}\psip_{p}
-i\textstyle{\frac{73}{18}}\brpsip_{B}\brgamma_{A}\psip_{C}\\
{}&-i\textstyle{\frac{5}{4}}\brrho\gamma_{ABC}\rho
+i\textstyle{\frac{5}{2}}\brrho\gamma_{BC}\psi_{A}
+i\textstyle{\frac{5}{4}}\brpsi^{\brp}\gamma_{ABC}\psi_{\brp}
-5i\brpsi_{B}\gamma_{A}\psi_{C}\,,
\ea
\label{GGG}
\ee
which  are designed to serve as  common  connections for all the equations of motion, see Appendix~\ref{secAppvarL}.\\

\noindent Under the  ${\cN=2}$ supersymmetry~(\ref{N2SUSY}),   disregarding  total derivatives, the Lagrangian transforms  concisely,
\be
\deltaS\cL_{\typeT}\seceq -\textstyle{\frac{1}{8}} e^{-2d}\brV^{A}{}_{\brq}\deltaS V_{Ap}\Tr\left(\gamma^{p}\tcF_{-}\brgamma^{\brq}\overline{\tcF_{-}}\,\right)\,.
\label{varLSUSY}
\ee
This verifies, to the full order in fermions, the {supersymmetric  invariance} of the type II SDFT action  modulo the self-duality~(\ref{SD}), see Appendix~\ref{secAppsusy} for  details.  For a nontrivial {consistency check}, the supersymmetric  variation  of the self-duality  relation~(\ref{SD}) is, to the full order {precisely}, closed by  the equations of motion for fermions, especially the   gravitinos (\ref{varL}),   
\be
\deltaS\tcF_{-}=-i\left(\tcD_{\brp}\rho+\gamma^{p}\tcD_{p}\psi_{\brp}-\gamma^{p}\cF\brgamma_{\brp}\psip_{p}\right)\brvarepsilonp\brgamma^{\brp}-i\gamma^{p}\varepsilon\left(\tcD^{\prime}_{p}\brrhop+\tcDp_{\brp}\brpsi^{\prime}_{p}\brgamma^{\brp}-\brpsi_{\brp}\gamma_{p}\cF\brgamma^{\brp}\right)\,.
\label{closed}
\ee
~\\

\section*{Unification}  
As stressed before, one of the characteristic features in our construction of ${\cN=2}$ ${D=10}$ SDFT is the usage of  the covariant  \textit{fundamental fields},  identified  in (\ref{FFC}).  However, the relation  to an ordinary  supergravity can  be   established   only after  we solve the defining algebraic relations of the DFT-vielbeins~(\ref{defV}) and parametrize the solution in terms of zehnbeins and $B$-field: Up to  $\Ott$ rotations and field redefinitions,  the generic solution reads~\cite{Jeon:2011cn,Jeon:2012kd}
\be
\ba{ll}
V_{Ap}=\textstyle{\frac{1}{\sqrt{2}}}{{\left(\ba{c} (e^{-1})_{p}{}^{\mu}\\(B+e)_{\nu p}\ea\right)}}\,,~
&~\brV_{A{\brp}}=\textstyle{\frac{1}{\sqrt{2}}}\left(\ba{c} (\bre^{-1})_{\brp}{}^{\mu}\\(B+\bre)_{\nu{\brp}}\ea\right)\,,
\ea
\label{Vform1}
\ee
where    $e_{\mu}{}^{p}$ and $\bre_{\nu}{}^{{\brp}}$ are two copies of  zehnbeins  which must constitute   a common   spacetime metric,   
\be
e_{\mu}{}^{p}e_{\nu}{}^{q}\eta_{pq}=-\bre_{\mu}{}^{{\brp}}\bre_{\nu}{}^{\brq}\breta_{\brp\brq}=g_{\mu\nu}\,.
\ee
We also set $B_{\mu p}=B_{\mu\nu}(e^{-1})_{p}{}^{\nu}$ and $B_{\mu\brp}=B_{\mu\nu}(\bre^{-1})_{{\brp}}{}^{\nu}$.  The pair of zehnbeins  directly reflects  the double local Lorentz groups, ${\Spint\times\oSpint}$.  ~It follows that  $(e^{-1}\bre)_{p}{}^{\brp}$ is a Lorentz rotation, 
 \be
(e^{-1}\bre)_{p}{}^{\brp}(e^{-1}\bre)_{q}{}^{\brq}\breta_{\brp\brq}=-\eta_{pq}\,,
\ee
and further that  there is a    spinorial representation of this  Lorentz rotation which relates\footnote{Since, as seen from  Table~\ref{TABindices}, our convention assumes the signature of $\,\eta_{pq}$ for $\Spint$ to be opposite to that of  $\,\breta_{\brp\brq}$ for  $\oSpint$, the spinorial representation, $S_{e}$, relates  $\brgamma^{\brp}$ to    $\gamma^{\eleven}\gamma^{p}$ rather than   $\gamma^{p}$. Note the  minus sign, 
\[\{\gamma^{\eleven}\gamma^{p},\gamma^{\eleven}\gamma^{q}\}=-2\eta^{pq}\,.\]}  $\brgamma^{\brp}$ to  $\gamma^{\eleven}\gamma^{p}$,
\be
S_{e}\brgamma^{\brp}S_{e}^{-1}=\gamma^{\eleven}\gamma^{p}(e^{-1}\bre)_{p}{}^{\brp}\,.
\label{LS}
\ee
\indent Now we may consider  `fixing'  the two  zehnbeins equal to each other, 
\be
{e_{\mu}{}^{p}\equiv\bre_{\mu}{}^{{\brp}}}\,,
\label{dfix}
\ee
using a  $\oPint$ local Lorentz rotation which effectively ``unwinds"  $(e^{-1}\bre)_{p}{}^{\brp}$ and $S_{e}$ such that they become   trivial \textit{i.e.}~identities.  This rotation   may, or may not,  flip the  chirality as 
\be
\signp~~\longrightarrow~~\det(e^{-1}\bre)\signp\,,
\ee 
 since (\ref{LS}) implies~\cite{Jeon:2012kd}
\be
S_{e}\brgamma^{\eleven}S_{e}^{-1}=-\det(e^{-1}\bre)\gamma^{\eleven}\,.
\ee
Namely, the chirality remains the same if  $\det(e^{-1}\bre)=+1$,  while  it  changes the sign   if  $\det(e^{-1}\bre)=-1$. 
Therefore, it depends on each specific background or each individual solution of the theory whether the chirality changes or not. That is to say, formulated   in terms of  the covariant  fields, \textit{i.e.} $V_{Ap}$, $\brV_{A\brp}$, $\cC^{\alpha}{}_{\bralpha}$, \textit{etc.}  the   ${\cN=2}$ ${D=10}$ SDFT is simply  a chiral theory  with respect to the pair of local Lorentz groups. All the possible chirality choices are  equivalent and hence  the theory is \textit{unique}. We  may safely put   ${\sign\equiv\signp\equiv+1}$ without loss of generality.  However, the theory contains two `types' of solutions.  All the solutions are  classified  into  two groups, 
\be
\ba{l}
\sign\signp\det(e^{-1}\bre)=+1~~~:~~~\mbox{type~IIA}\,,\\
\sign\signp\det(e^{-1}\bre)=-1~~~:~~~\mbox{type~IIB}\,.
\ea
\label{classify}
\ee
Conversely,  making full use of  the above ${\oPint}$ rotation,   any    solution in type IIA and  type IIB supergravities  can be mapped  to  a solution of ${\cN=2}$ ${D=10}$  SDFT of fixed chirality  \textit{e.g.}   ${\sign\equiv\signp\equiv+1}$.  
\textit{The single unique ${\cN=2}$ ${D=10}$ SDFT unifies  type IIA and IIB supergravities.}

\section*{Comments}  
After the fixing, ${e_{\mu}{}^{p}\equiv\bre_{\mu}{}^{{\brp}}}$ (\ref{dfix}), the pair of local Lorentz groups, $\Spint\times\oSpint$,  is   broken  to  its diagonal subgroup, $\Spint_{{\scriptscriptstyle{\mathbf{{D}}}}}$, which  acts on both $\Pint$  and $\oPint$  indices simultaneously.   This allows  us to expand  $\cC^{\alpha}{}_{\bralpha}$ in terms of odd (type IIA)  or even (type IIB) $p$-forms~\cite{Jeon:2012kd}, and   eventually reduces the ${\cN=2}$ ${D=10}$ SDFT  to the  so-called    `democratic supergravity' formulated,  up to   quadratic order in fermions,   in \cite{Bergshoeff:2001pv}  (\textit{c.f.~}\cite{Townsend:1995gp,Green:1996bh}). \\

\noindent The diagonal  ``gauge"  fixing (\ref{dfix})  inevitably  modifies the $\Ott$ T-duality transformation rule to call for a  compensating $\oPint$ local Lorentz  rotation~\cite{Jeon:2012kd}, such that the  fermions and the R-R sector are \textit{no longer} $\Ott$ singlets. In particular, the R-R sector can be  mapped to the  $\Ott$ spinor  in \cite{Fukuma:1999jt,Hassan:1999mm,Hohm:2011zr,Hohm:2011dv}. Moreover,  the modified $\Ott$ T-duality  transformation,  or more precisely the  compensating $\oPint$ local Lorentz  rotation,  may    flip the chirality of the theory, resulting in the usual exchange of  IIA and IIB. \\

\noindent However, \textit{a priori} T-duality is not a Noether symmetry. It becomes so only if it acts on an isometry direction. Hence,  as is well known,   within the supergravity setup  the equivalence between IIA and IIB can be established  only when the background admits  an isometry. This is compared to the     `background independent'   unification of the two supergravities   by  ${\cN=2}$ ${D=10}$ SDFT,  discussed in this work. \\

\noindent Turning off  both the primed fermions and the R-R sector   truncates   the  $\cN=2$  SDFT to the previously constructed  $\cN=1$ SDFT~\cite{Jeon:2011sq},   to the full order in fermions consistently~(\textit{c.f.~}\cite{Hohm:2011nu}). The uplift  of    type II SDFT to $\cM$-theory,  or the extension of $\Ott$ T-duality  to  $E_{11}$  U-duality,  remains as a challenging future work, \textit{c.f.}~\cite{Berman:2010is,Thompson:2011uw,Berman:2011cg,Berman:2011jh,West:2011mm,West:2001as,West:2010ev,Rocen:2010bk,Coimbra:2011ky,Bandos:1997gd}.  \\

\noindent The Appendices  contain  some details of the computations for  (\ref{varL}) and (\ref{varLSUSY}).  \\
 
\noindent{{\textbf{Acknowledgements.}}   The work was supported by the National Research Foundation of Korea and  the Ministry of Education, Science and Technology with the Grant No.  2005-0049409 (CQUeST), No.  2010-0002980, No. 2012R1A2A2A02046739 and  No. 2012R1A6A3A03040350.

 \newpage

\appendix

\renewcommand{\theequation}{\thesection.\arabic{equation}}
\@addtoreset{equation}{section} \makeatother
\setlength{\jot}{9pt}                 
\renewcommand{\arraystretch}{1.8}

\begin{center}
\bf{{APPENDICES}}
\end{center}

\section{Variation of the Lagrangian under arbitrary transformations of fields\label{secAppvarL}}
For arbitrary variations of  fields,  the  identities below    hold  either   strictly (`$\,=\,$')\,   or   up to  total derivatives  and the section condition (`$\,\seceq\,$').\\

For the double-vielbein,  generic (torsionful) connection  and  curvature,
\be
\ba{ll}
\delta V_{Ap}=\brP_{A}{}^{B}\delta V_{Bp}+V_{A}{}^{q}\delta V_{B[p}V^{B}{}_{q]}\,,~&~~~
\delta\brV_{A\brp}=P_{A}{}^{B}\delta\brV_{B\brp}+\brV_{A}{}^{\brq}\delta\brV_{B[\brp}\brV^{B}{}_{\brq]}\,,\\
\delta\Phi_{Apq}=\cD_{A}(V^{B}{}_{p}\delta V_{Bq})+V^{B}{}_{p}V^{C}{}_{q}\delta\Gamma_{ABC}\,,~&~~~
\delta\brPhi_{A\brp\brq}=\cD_{A}(\brV^{B}{}_{\brp}\delta \brV_{B\brq})+\brV^{B}{}_{\brp}\brV^{C}{}_{\brq}\delta\Gamma_{ABC}\,,\\
\multicolumn{2}{c}{\delta S_{ABCD}=\cD_{[A}\delta\Gamma_{B]CD}+\cD_{[C}\delta\Gamma_{D]AB}
-\textstyle{\frac{3}{2}}\Gamma_{[ABE]}\delta\Gamma^{E}{}_{CD}
-\textstyle{\frac{3}{2}}\Gamma_{[CDE]}\delta\Gamma^{E}{}_{AB}\,.}
\ea
\label{useful2}
\ee
Further with the fermions,
\be
\ba{l}
\delta V_{Ap}\brpsi_{\brp}\gamma^{Ap}\gamma^{abc}\psi^{\brp}\,\brrho\gamma_{abc}\rho
+\delta V_{Ap}\brrho\gamma^{Ap}\gamma^{abc}\rho\,\brpsi_{\brp}\gamma_{abc}\psi^{\brp}=0\,,\\
\delta\brV_{A\brp}\brpsi^{\prime}_{p}\brgamma^{A\brp}\brgamma^{\bra\brb\brc}\psi^{\prime p}\,\brrhop\brgamma_{\bra\brb\brc}\rhop+\delta\brV_{A\brp}\brrhop\brgamma^{A\brp}\brgamma^{\bra\brb\brc}\rhop\,\brpsip_{p}\brgamma_{\bra\brb\brc}\psi^{\prime p}=0\,.
\ea
\ee

For the NS-NS sector of the Lagrangian,   
\be
\delta\!\left[\textstyle{\frac{1}{8}}(P^{AB}P^{CD}-\brP^{AB}\brP^{CD})S_{ACBD}\right]\seceq
\half\delta V^{Ap}\brV_{A}{}^{\brq}S_{p\brq}-\textstyle{\frac{3}{8}}\delta\Gamma_{ABC}(P^{B}{}_{D}P^{C}{}_{E}-\brP^{B}{}_{D}\brP^{C}{}_{E})\Gamma^{[ADE]}\,.
\ee

For   an arbitrary bi-fundamental quantity,  $\cM^{\alpha}{}_{\bralpha}$,  with  the charge conjugation,  $\bar{\cM}=\brC_{+}^{-1}\cM^{T}C_{+}$,
\be
\ba{ll}
e^{-2d}\Tr(\delta\cF\bar{\cM})\seceq&e^{-2d\,}\delta d\,\Tr(\cF\bar{\cM})\\
{}&\!\!\!\!\!\!\!+e^{-2d}\Tr\Big[\!\left(-\delta\cC+\cC\delta d
-\quarter \delta V_{Ap}\gamma^{Ap}\cC-\half\brV^{A}{}_{\brp}\delta V_{Aq}\gamma^{\eleven}\gamma^{q}\cC\brgamma^{\brp}+\quarter\delta\brV_{A\brp}\cC\brgamma^{A\brp}
\right)\overline{\cDo_{-}\cM}\\
{}&\quad\quad\quad\quad\quad+\left(-\quarter \delta V_{Ap}\gamma^{Ap}\cF-\half\brV^{A}{}_{\brp}\delta V_{Aq}\gamma^{\eleven}\gamma^{q}\cF\brgamma^{\brp}+\quarter\delta\brV_{A\brp}\cF\brgamma^{A\brp}\right)\bar{\cM}\,\Big]\,.
\ea
\ee
Hence,  for the R-R sector  of the Lagrangian, we obtain
\be
\ba{ll}
\multicolumn{2}{l}{
\delta\!\left[e^{-2d}\left(\half\cF^{\alpha\bralpha}\cF_{\alpha\bralpha}-i\brrho\cF\rhop+i\brpsi_{\brp}\gamma_{q}\cF\brgamma^{\brp}\psip{}^{q}\right)\right]}\\
~\seceq&e^{-2d\,}\delta d\left(i\brrho\cF\rhop-i\brpsi_{\brp}\gamma_{q}\cF\brgamma^{\brp}\psip{}^{q}
\right) \\
{}&+e^{-2d}\left[
-i\left(\delta\brrho-\textstyle{\frac{1}{4}}\delta V_{Bq}\brrho\gamma^{Bq}\right)
\cF\rhop+i\left(\delta\brpsi_{\brp}-\delta\brV^{B}{}_{\brp}\brpsi_{B}
-\textstyle{\frac{1}{4}}\delta V_{Bq}\brpsi_{\brp}\gamma^{Bq}\right)\gamma^{q}\cF\brgamma^{\brp}\psi^{\prime}_{q}\right]\\
{}&+e^{-2d}\left[
+i\left(\delta\brrhop-\textstyle{\frac{1}{4}}\delta\brV_{B\brq}\brrhop\brgamma^{B\brq}\right)\brcF\rho
-i\left(\delta\brpsi^{\prime}_{p}-\delta V^{B}{}_{p}\brpsi^{\prime}_{B}
-\textstyle{\frac{1}{4}}\delta\brV_{B\brq}\brpsi^{\prime}_{p}\brgamma^{B\brq}\right)\brgamma^{\brq}\brcF\gamma^{p}\psi_{\brq}\right]\\
{}&+\half e^{-2d}\delta V^{Ap}\brV_{A}{}^{\brq\,}\Tr\!\left[\gamma^{\eleven}\left(\cF-i\rho\brrhop+i\gamma^{r}\psi_{\brs}\brpsi^{\prime}_{r}\brgamma^{\brs}\right)\brgamma_{\brq}\brcF\gamma_{p}\right]\\
{}&-e^{-2d}\left(\delta\cC -\cC\delta d 
+\quarter \delta V_{Ap}\gamma^{Ap}\cC+\half\delta V_{Ap}\gamma^{\eleven}\gamma^{p}\cC\brgamma^{A}-\quarter\delta\brV_{A\brp}\cC\brgamma^{A\brp}
\right)^{\alpha\bralpha}\\
{}&\quad\quad\quad\quad\quad\quad\quad\quad\quad\times\left[\cDo_{-}\left(\cF-i\rho\brrhop+i\gamma^{r}\psi_{\brs}\brpsi^{\prime}_{r}\brgamma^{\brs}\right)\right]_{\alpha\bralpha}\,.
\ea
\label{RRvar}
\ee

For the    fermionic  kinetic terms,  from (\ref{useful2}), we have
\be
\ba{ll}
\multicolumn{2}{l}{e^{-2d}\delta\!\left(i\half\brrho \gamma^{p}\cDs_{p}\rho
-i\brpsi^{\brp}\cDs_{\brp}\rho-i\half\brpsi^{\brp}\gamma^{q}\cDs_{q}\psi_{\brp}
-i\half\brrhop \brgamma^{\brp}\cDps_{\brp}\rhop
+i\brpsip{}^{p}\cDps_{p}\rhop+i\half\brpsip{}^{p}\brgamma^{\brq}\cDps_{\brq}\psi^{\prime}_{p}\right)}\\
\seceq&
i\half e^{-2d}\delta V^{Bp}\brV_{B}{}^{\brq}\left(
\brrho\gamma_{p}\cDs_{\brq}\rho+2\brpsi_{\brq}\cDs_{p}\rho-\brpsi^{\brp}\gamma_{p}\cDs_{\brq}\psi_{\brp}
+\brrhop\brgamma_{\brq}\cDps_{p}\rhop
+2\brpsi^{\prime}_{p}\cDps_{\brq}\rhop-\brpsip{}^{q}\brgamma_{\brq}\cDps_{p}\psi^{\prime}_{q}\right)\\
{}&+ie^{-2d}\left(\delta\brrho-\textstyle{\frac{1}{4}}\delta V_{Bq}\brrho\gamma^{Bq}\right)
\left(\gamma^{p}\cDs_{p}\rho-\cDs_{\brp}\psi^{\brp}\right)\\
{}&-ie^{-2d}\left(\delta\brpsi^{\brp}-\delta\brV^{B\brp}\brpsi_{B}
-\textstyle{\frac{1}{4}}\delta V_{Bq}\brpsi^{\brp}\gamma^{Bq}\right)\left(\cDs_{\brp}\rho+\gamma^{p}\cDs_{p}\psi_{\brp}\right)\\
{}&-ie^{-2d}\left(\delta\brrhop-\textstyle{\frac{1}{4}}\delta\brV_{B\brq}\brrhop\brgamma^{B\brq}\right)
\left(\brgamma^{\brp}\cDps_{\brp}\rhop-\cDps_{p}\psip{}^{p}\right)\\
{}&+ie^{-2d}\left(\delta\brpsip{}^{p}-\delta V^{Bp}\brpsi^{\prime}_{B}
-\textstyle{\frac{1}{4}}\delta\brV_{B\brq}\brpsip{}^{p}\brgamma^{B\brq}\right)\left(\cDps_{p}\rhop+\brgamma^{\brp}\cDps_{\brp}\psi^{\prime}_{p}\right)\\
{}&+ie^{-2d}\delta\Gammas_{ABC}\left(\textstyle{\frac{1}{8}}
\brrho\gamma^{ABC}\rho-\quarter\brpsi^{A}\gamma^{BC}\rho-\textstyle{\frac{1}{8}}\brpsi^{\brp}\gamma^{ABC}\psi_{\brp}-\half\brpsi^{B}\gamma^{A}\psi^{C}\right)\\
{}&-ie^{-2d}\delta\Gammaps_{ABC}\left(\textstyle{\frac{1}{8}}
\brrhop\brgamma^{ABC}\rhop-\quarter\brpsip{}^{A}\brgamma^{BC}\rhop-\textstyle{\frac{1}{8}}\brpsip{}^{p}\brgamma^{ABC}\psi_{p}-\half\brpsip{}^{B}\brgamma^{A}\psip{}^{C}\right)\,.
\ea
\label{KF}
\ee
Here we let the connections  assume the following  generic forms:
\be
\ba{ll}
\Gammas_{ABC}=&\Gamma_{ABC}+a_{1}\brrho\gamma_{ABC}\rho+a_{2}\brrho\gamma_{BC}\psi_{A}+a_{3}\brpsi_{\brp}\gamma_{ABC}\psi^{\brp}+a_{4}\brpsi_{B}\gamma_{A}\psi_{C}\\
{}&+\aap_{1}\brrhop\brgamma_{ABC}\rhop+\aap_{2}\brrhop\brgamma_{BC}\psip_{A}+\aap_{3}\brpsip_{p}\brgamma_{ABC}\psi^{\prime p}+\aap_{4}\brpsip_{B}\brgamma_{A}\psip_{C}\,,\\
\Gammaps_{ABC}=&\Gamma_{ABC}+a_{1}\brrhop\brgamma_{ABC}\rhop+a_{2}\brrhop\brgamma_{BC}\psip_{A}+a_{3}\brpsip_{p}\brgamma_{ABC}\psi^{\prime p}+a_{4}\brpsip_{B}\brgamma_{A}\psip_{C}\\
{}&+\aap_{1}\brrho\gamma_{ABC}\rho+\aap_{2}\brrho\gamma_{BC}\psi_{A}+\aap_{3}\brpsi_{\brp}\gamma_{ABC}\psi^{\brp}+\aap_{4}\brpsi_{B}\gamma_{A}\psi_{C}\,.
\ea
\label{Gammasps}
\ee
It is easy to check  that,   $\aap_{1}$ and $\aap_{3}$ decouple from the  fermionic kinetic terms~(\ref{KF}), and only the linear combination, $\aap_{2}-\half\aap_{4\,}$ alone is relevant among  the four primed coefficients, $\{\aap_{1}, \aap_{2}, \aap_{3}, \aap_{4}\}$. Without loss of generality,  henceforth we   put
\be
\aap_{1}=\aap_{3}=0\,.
\label{aapzero}
\ee
We proceed to compute  the variations of $\Gammas_{ABC}$ and $\Gammaps_{ABC}$ (\ref{Gammasps}), for which  we first note 
\be
\ba{l}
\delta\!\left(\aap_{1}\brrhop\brgamma_{ABC}\rhop+\aap_{2}\brrhop\brgamma_{BC}\psi^{\prime}_{A}+\aap_{3}\brpsi^{\prime}_{p}\brgamma_{ABC}\psi^{\prime p}+\aap_{4}\brpsi^{\prime}_{B}\brgamma_{A}\psi^{\prime}_{C}\right)\\
\quad\quad\times
\left(\textstyle{\frac{1}{8}}
\brrho\gamma^{ABC}\rho-\quarter\brpsi^{A}\gamma^{BC}\rho-\textstyle{\frac{1}{8}}\brpsi^{\brp}\gamma^{ABC}\psi_{\brp}-\half\brpsi^{B}\gamma^{A}\psi^{C}\right)\\
=\delta V^{Ap}\brV_{A}{}^{\brp}\left(
\half \aap_{1}\brrhop\brgamma_{\brp\brq\brr}\rhop\brpsi^{\brq}\gamma_{p}\psi^{\brr}+
\half \aap_{3}\brpsi^{\prime}_{q}\brgamma_{\brp\brq\brr}\psi^{\prime q}\brpsi^{\brq}\gamma_{p}\psi^{\brr}-\textstyle{\frac{1}{8}}\aap_{4}\brpsi^{\prime q}\brgamma_{\brp}\psi^{\prime r}\brrho\gamma_{pqr}\rho
+\textstyle{\frac{1}{8}}\aap_{4}\brpsi^{\prime q}\brgamma_{\brp}\psi^{\prime r}\brpsi_{\brq}\gamma_{pqr}\psi^{\brq}
\right)\\
\quad-\half \aap_{2}\left(\delta\brrhop-\textstyle{\frac{1}{4}}\delta \brV_{B\brp}\brrhop\brgamma^{B\brp}\right)\brgamma^{\brq\brr}\psi^{\prime p}\brpsi_{\brq}\gamma_{p}\psi_{\brr}-\half \aap_{2}\left(\delta\brpsi^{\prime p}-\delta V^{Bp}\brpsip_{B}
-\textstyle{\frac{1}{4}}\delta\brV_{B\brq}\brpsi^{\prime p}\brgamma^{B\brq}\right)\brgamma^{\brr\brs}\rho^{\prime}\brpsi_{\brr}\gamma_{p}\psi_{\brs}\\
\quad-\half \aap_{4}\left(\delta\brpsi^{\prime p}-\delta V^{Bp}\brpsip_{B}
-\textstyle{\frac{1}{4}}\delta\brV_{B\brq}\brpsi^{\prime p}\brgamma^{B\brq}\right)\brgamma^{\brp}\psi^{\prime q}
\brrho\gamma_{pq}\psi_{\brp}\,.
\ea
\ee
Yet, with (\ref{aapzero}) taken, we  just  need
\be
\ba{l}
\delta\!\left(a_{1}\brrho\gamma_{ABC}\rho+a_{2}\brrho\gamma_{BC}\psi_{A}+a_{3}\brpsi_{\brp}\gamma_{ABC}\psi^{\brp}+a_{4}\brpsi_{B}\gamma_{A}\psi_{C}+\aap_{2}\brrhop\brgamma_{BC}\psi^{\prime}_{A}+\aap_{4}\brpsip_{B}\brgamma_{A}\psip_{C}\right)\\
\quad\quad\quad\times
\left(\textstyle{\frac{1}{8}}
\brrho\gamma^{ABC}\rho-\quarter\brpsi^{A}\gamma^{BC}\rho-\textstyle{\frac{1}{8}}\brpsi^{\brp}\gamma^{ABC}\psi_{\brp}-\half\brpsi^{B}\gamma^{A}\psi^{C}\right)\\
=\delta V^{Ap}\brV_{A}{}^{\brp}\left[-(\quarter a_{1}+\textstyle{\frac{1}{8}}a_{2})\brrho\gamma^{st}\psi_{\brp}\brrho\gamma_{pst}\rho
+(\textstyle{\frac{1}{8}}a_{2}-\quarter a_{3})\brrho\gamma^{st}\psi_{\brp}\brpsi_{\brq}\gamma_{pst}\psi^{\brq}\right.\\
~\quad\quad\quad\quad\quad\quad\left.
-\textstyle{\frac{1}{8}}\aap_{4}\brpsi^{\prime q}\brgamma_{\brp}\psi^{\prime r}\brrho\gamma_{pqr}\rho
+\textstyle{\frac{1}{8}}\aap_{4}\brpsi^{\prime q}\brgamma_{\brp}\psi^{\prime r}\brpsi_{\brq}\gamma_{pqr}\psi^{\brq}
\right]\\
~\quad+\left(\delta\brrho-\textstyle{\frac{1}{4}}\delta V_{Bp}\brrho\gamma^{Bp}\right)
\left[\quarter\gamma_{rst}\rho(-a_{1}+\textstyle{\frac{1}{16}}a_{2})\brpsi_{\brp}\gamma^{rst}\psi^{\brp}\right]\\
~\quad+\left(\delta\brpsi^{\brp}-\delta\brV^{B\brp}\brpsi_{B}
-\textstyle{\frac{1}{4}}\delta V_{Bp}\brpsi^{\brp}\gamma^{Bp}\right)
\left[\quarter\gamma_{rst}\psi_{\brp}\left((\textstyle{\frac{1}{16}}a_{2}+a_{3})\brrho\gamma^{rst}\rho
-(a_{3}+\textstyle{\frac{1}{6}}a_{4})\brpsi_{\brq}\gamma^{rst}\psi^{\brq}\right)\right]\\
~\quad-\half \aap_{2}\left(\delta\brrhop-\textstyle{\frac{1}{4}}\delta \brV_{B\brp}\brrhop\brgamma^{B\brp}\right)\brgamma^{\brq\brr}\psi^{\prime p}\brpsi_{\brq}\gamma_{p}\psi_{\brr}\\
~\quad-\half \aap_{2}\left(\delta\brpsi^{\prime p}-\delta V^{Bp}\brpsip_{B}
-\textstyle{\frac{1}{4}}\delta\brV_{B\brq}\brpsi^{\prime p}\brgamma^{B\brq}\right)\brgamma^{\brr\brs}\rho^{\prime}\brpsi_{\brr}\gamma_{p}\psi_{\brs}\\
~\quad-\half \aap_{4}\left(\delta\brpsi^{\prime p}-\delta V^{Bp}\brpsip_{B}
-\textstyle{\frac{1}{4}}\delta\brV_{B\brq}\brpsi^{\prime p}\brgamma^{B\brq}\right)\brgamma^{\brp}\psi^{\prime q}
\brrho\gamma_{pq}\psi_{\brp}\,,
\ea
\ee
and  similarly 
\be
\ba{l}
\delta\!\left(a_{1}\brrhop\brgamma_{ABC}\rhop+a_{2}\brrhop\brgamma_{BC}\psip_{A}+a_{3}\brpsip_{p}\brgamma_{ABC}\psi^{\prime p}+a_{4}\brpsip_{B}\brgamma_{A}\psip_{C}+\aap_{2}\brrho\gamma_{BC}\psi_{A}+\aap_{4}\brpsi_{B}\gamma_{A}\psi_{C}\right)\\
\quad\quad\quad\times
\left(\textstyle{\frac{1}{8}}
\brrhop\brgamma^{ABC}\rhop-\quarter\brpsip{}^{A}\brgamma^{BC}\rhop-\textstyle{\frac{1}{8}}\brpsip{}^{p}\brgamma^{ABC}\psi_{p}-\half\brpsip{}^{B}\brgamma^{A}\psip{}^{C}\right)\\
=-\delta V^{Ap}\brV_{A}{}^{\brp}\left[-(\quarter a_{1}+\textstyle{\frac{1}{8}}a_{2})\brrhop\brgamma^{\brs\brt}\psip_{p}\brrhop\brgamma_{\brp\brs\brt}\rhop+
(\textstyle{\frac{1}{8}}a_{2}-\quarter a_{3})\brrhop\brgamma^{\brs\brt}\psip_{p}\brpsip_{q}\brgamma_{\brp\brs\brt}\psi^{\prime q}\right.\\
~\quad\quad\quad\quad\quad\quad\quad\left.
-\textstyle{\frac{1}{8}}\aap_{4}\brpsi^{\brq}\gamma_{p}\psi^{\brr}\brrho^{\prime}\brgamma_{\brp\brq\brr}\rho^{\prime}
+\textstyle{\frac{1}{8}}\aap_{4}\brpsi^{\brq}\gamma_{p}\psi^{\brr}\brpsi^{\prime}_{q}\brgamma_{\brp\brq\brr}\psi^{\prime q}
\right]\\
~\quad+\left(\delta\brrhop-\textstyle{\frac{1}{4}}\delta \brV_{B\brp}\brrhop\brgamma^{B\brp}\right)
\left[\quarter\brgamma_{\brr\brs\brt}\rhop(-a_{1}+\textstyle{\frac{1}{16}}a_{2})\brpsip_{p}\brgamma^{\brr\brs\brt}\psi^{\prime p}\right]\\
~\quad+\left(\delta\brpsi^{\prime p}-\delta V^{Bp}\brpsip_{B}
-\textstyle{\frac{1}{4}}\delta\brV_{B\brq}\brpsi^{\prime p}\brgamma^{B\brq}\right)
\left[\quarter\brgamma_{\brr\brs\brt}\psip_{p}\left((\textstyle{\frac{1}{16}}a_{2}+a_{3})\brrhop\brgamma^{\brr\brs\brt}\rhop
-(a_{3}+\textstyle{\frac{1}{6}}a_{4})\brpsip_{q}\brgamma^{\brr\brs\brt}\psi^{\prime q}\right)\right]\\
~\quad-\half \aap_{2}\left(\delta\brrho-\textstyle{\frac{1}{4}}\delta V_{Bp}\brrho\gamma^{Bp}\right)\gamma^{qr}\psi^{\brp}\brpsi^{\prime}_{q}\brgamma_{\brp}\psi^{\prime}_{r}\\
~\quad-\half \aap_{2}\left(\delta\brpsi^{\brp}-\delta\brV^{B\brp}\brpsi_{B}
-\textstyle{\frac{1}{4}}\delta V_{Bp}\brpsi^{\brp}\gamma^{Bp}\right)\gamma^{qr}\rho{}
\brpsi^{\prime}_{q}\brgamma_{\brp}\psi^{\prime}_{r}\\
~\quad-\half \aap_{4}\left(\delta\brpsi^{\brp}-\delta \brV^{B\brp}\brpsi_{B}
-\textstyle{\frac{1}{4}}\delta V_{Bp}\brpsi^{\brp}\gamma^{Bp}\right)\gamma^{q}\psi^{\brq}
\brrhop\brgamma_{\brp\brq}\psi^{\prime}_{q}\,.
\ea
\ee
The variation of the fermionic kinetic terms  (\ref{KF})  now assumes the desired expression:
\be
\ba{ll}
\multicolumn{2}{l}{e^{-2d}\delta\!\left(i\half\brrho \gamma^{p}\cDs_{p}\rho
-i\brpsi^{\brp}\cDs_{\brp}\rho-i\half\brpsi^{\brp}\gamma^{q}\cDs_{q}\psi_{\brp}
-i\half\brrhop \brgamma^{\brp}\cDps_{\brp}\rhop
+i\brpsip{}^{p}\cDps_{p}\rhop+i\half\brpsip{}^{p}\brgamma^{\brq}\cDps_{\brq}\psi^{\prime}_{p}\right)}\\
\seceq&
i\half e^{-2d}\delta V^{Bp}\brV_{B}{}^{\brq}\left(
\brrho\gamma_{p}\cDsh_{\brq}\rho+2\brpsi_{\brq}\tcD_{p}\rho-\brpsi^{\brp}\gamma_{p}\cDfl_{\brq}\psi_{\brp}+\brrhop\brgamma_{\brq}\cDpsh_{p}\rhop+2\brpsi^{\prime}_{p}\tcDp_{\brq}\rhop-\brpsip{}^{q}\brgamma_{\brq}\cDpfl_{p}\psi^{\prime}_{q}\right)\\
{}&+ie^{-2d}\left(\delta\brrho-\textstyle{\frac{1}{4}}\delta V_{Bq}\brrho\gamma^{Bq}\right)
\left(\gamma^{p}\cDsh_{p}\rho-\tcD_{\brp}\psi^{\brp}\right)\\
{}&-ie^{-2d}\left(\delta\brpsi^{\brp}-\delta\brV^{B\brp}\brpsi_{B}
-\textstyle{\frac{1}{4}}\delta V_{Bq}\brpsi^{\brp}\gamma^{Bq}\right)
\left(\tcD_{\brp}\rho+\gamma^{p}\cDfl_{p}\psi_{\brp}\right)\\
{}&-ie^{-2d}\left(\delta\brrhop-\textstyle{\frac{1}{4}}\delta\brV_{B\brq}\brrhop\brgamma^{B\brq}\right)\left(\brgamma^{\brp}\cDpsh_{\brp}\rhop-\tcDp_{p}\psip{}^{p}\right)\\
{}&+ie^{-2d}\left(\delta\brpsip{}^{p}-\delta V^{Bp}\brpsi^{\prime}_{B}
-\textstyle{\frac{1}{4}}\delta\brV_{B\brq}\brpsip{}^{p}\brgamma^{B\brq}\right)
\left(\tcDp_{p}\rhop+\brgamma^{\brp}\cDpfl_{\brp}\psi^{\prime}_{p}\right)\\
{}&+ie^{-2d}\delta\Gamma_{ABC}\left(\textstyle{\frac{1}{8}}
\brrho\gamma^{ABC}\rho-\quarter\brpsi^{A}\gamma^{BC}\rho-\textstyle{\frac{1}{8}}\brpsi^{\brp}\gamma^{ABC}\psi_{\brp}-\half\brpsi^{B}\gamma^{A}\psi^{C}\right.\\
{}&\quad\quad\quad\quad\quad\quad~~\left.-\textstyle{\frac{1}{8}}
\brrhop\brgamma^{ABC}\rhop+\quarter\brpsip{}^{A}\brgamma^{BC}\rhop+\textstyle{\frac{1}{8}}\brpsip{}^{p}\brgamma^{ABC}\psi_{p}+\half\brpsip{}^{B}\brgamma^{A}\psip{}^{C}\right)\,,
\ea
\label{KF2}
\ee
and  the Lagrangian transforms  up to total derivatives as
\be
\ba{ll}
\delta\cL_{\typeT}\seceq&-2\delta d\times\cL_{\typeT}\\
{}&+\delta\Gamma_{ABC}\times 0\\
{}&+\half e^{-2d} \delta V^{Bp}\brV_{B}{}^{\brq}\left[\tilde{S}_{p\brq}
+\Tr(\cF\brgamma_{\brq}\brcF\gamma_{p})
\right]\\
{}&-ie^{-2d}\left(\delta\brpsi^{\brp}-\delta\brV^{B\brp} \brpsi_{B}
-\textstyle{\frac{1}{4}}\delta V_{Bq}\brpsi^{\brp}\gamma^{Bq}\right)
\left(\tcD_{\brp}\rho+\gamma^{p}\cDfl_{p}\psi_{\brp}-\gamma^{p}\cF\brgamma_{\brp}\psip_{p}\right)\\
{}&+ie^{-2d}\left(\delta\brrho-\textstyle{\frac{1}{4}}\delta V_{Bq}\brrho\gamma^{Bq}\right)
\left(\gamma^{p}\cDsh_{p}\rho-\tcD_{\brp}\psi^{\brp}-\cF\rho^{\prime}\right)\\
{}&+ie^{-2d}\left(\delta\brpsi^{\prime p}-\delta V^{Bp}\brpsip_{B}
-\textstyle{\frac{1}{4}}\delta \brV_{B\brq}\brpsi^{\prime p}\brgamma^{B\brq}\right)
\left(\tcDp_{p}\rhop+\brgamma^{\brp}\cDpfl_{\brp}\psip_{p}-\brgamma^{\brp}\brcF\gamma_{p}\psi_{\brp}\right)\\
{}&-ie^{-2d}\left(\delta\brrhop-\textstyle{\frac{1}{4}}\delta \brV_{B\brq}\brrhop\brgamma^{B\brq}\right)
\left(\brgamma^{\brp}\cDpsh_{\brp}\rhop-\tcDp_{p}\psi^{\prime p}-\brcF\rho\right)\\
{}&+e^{-2d}\Tr\!\left[\tcF_{-}\left(\delta d\brcF-\half 
\delta V^{Ap}\brV_{A}{}^{\brq}\brgamma_{\brq}\brcF\gamma_{p}\right)-\cDo_{-}\tcF_{-}\overline{\widetilde{\delta\cC}} \right]\,.
\ea
\label{opf}
\ee
Here we set  generically
\be
\tilde{S}_{p\brq}:=S_{p\brq}+2i\brpsi_{\brq}\tcD_{p}\rho-i\brpsi^{\brp}\gamma_{p}\cDfl_{\brq}\psi_{\brp}+2i\brpsip_{p}\tcDp_{\brq}\rhop-i\brpsi^{\prime q}\brgamma_{\brq}\cDpfl_{p}\psip_{q}+i\brrho\gamma_{p}\cDsh_{\brq}\rho
+i\brrhop\brgamma_{\brq}\cDpsh_{p}\rhop\,,
\ee
and 
\be
\ba{ll}
\Gammash_{ABC}=&\Gammas_{ABC}+b_{1}\brrho\gamma_{ABC}\rho+b_{2}\brrho\gamma_{BC}\psi_{A}+b_{3}\brpsi_{\brp}\gamma_{ABC}\psi^{\brp}+b_{4}\brpsi_{B}\gamma_{A}\psi_{C}\\
{}&+\bbp_{1}\brrhop\brgamma_{ABC}\rhop+\bbp_{2}\brrhop\brgamma_{BC}\psip_{A}+\bbp_{3}\brpsip_{p}\brgamma_{ABC}\psi^{\prime p}+\bbp_{4}\brpsip_{B}\brgamma_{A}\psip_{C}\,,\\
\Gammafl_{ABC}=&\Gammas_{ABC}+c_{1}\brrho\gamma_{ABC}\rho+c_{2}\brrho\gamma_{BC}\psi_{A}+c_{3}\brpsi_{\brp}\gamma_{ABC}\psi^{\brp}+c_{4}\brpsi_{B}\gamma_{A}\psi_{C}\\
{}&+\ccp_{1}\brrhop\brgamma_{ABC}\rhop+\ccp_{2}\brrhop\brgamma_{BC}\psip_{A}+\ccp_{3}\brpsip_{p}\brgamma_{ABC}\psi^{\prime p}+\ccp_{4}\brpsip_{B}\brgamma_{A}\psip_{C}\,,\\
\tGamma_{ABC}=&\Gammas_{ABC}+d_{1}\brrho\gamma_{ABC}\rho+d_{2}\brrho\gamma_{BC}\psi_{A}+d_{3}\brpsi_{\brp}\gamma_{ABC}\psi^{\brp}+d_{4}\brpsi_{B}\gamma_{A}\psi_{C}\\
{}&+\ddp_{1}\brrhop\brgamma_{ABC}\rhop+\ddp_{2}\brrhop\brgamma_{BC}\psip_{A}+\ddp_{3}\brpsip_{p}\brgamma_{ABC}\psi^{\prime p}+\ddp_{4}\brpsip_{B}\brgamma_{A}\psip_{C}\,,
\ea
\ee
\be
\ba{ll}
\Gammapsh_{ABC}=&\Gammaps_{ABC}+b_{1}\brrhop\brgamma_{ABC}\rhop+b_{2}\brrhop\brgamma_{BC}\psip_{A}+b_{3}\brpsip_{p}\brgamma_{ABC}\psi^{\prime p}+b_{4}\brpsip_{B}\brgamma_{A}\psip_{C}\\
{}&+\bbp_{1}\brrho\gamma_{ABC}\rho+\bbp_{2}\brrho\gamma_{BC}\psi_{A}+\bbp_{3}\brpsi_{\brp}\gamma_{ABC}\psi^{\brp}+\bbp_{4}\brpsi_{B}\gamma_{A}\psi_{C}\,,\\
\Gammapfl_{ABC}=&\Gammaps_{ABC}+c_{1}\brrhop\brgamma_{ABC}\rhop+c_{2}\brrhop\brgamma_{BC}\psip_{A}+c_{3}\brpsip_{p}\brgamma_{ABC}\psi^{\prime p}+c_{4}\brpsip_{B}\brgamma_{A}\psip_{C}\\
{}&+\ccp_{1}\brrho\gamma_{ABC}\rho+\ccp_{2}\brrho\gamma_{BC}\psi_{A}+\ccp_{3}\brpsi_{\brp}\gamma_{ABC}\psi^{\brp}+\ccp_{4}\brpsi_{B}\gamma_{A}\psi_{C}\,,\\
\tGammap_{ABC}=&\Gammaps_{ABC}+d_{1}\brrhop\brgamma_{ABC}\rhop+d_{2}\brrhop\brgamma_{BC}\psip_{A}+d_{3}\brpsip_{p}\brgamma_{ABC}\psi^{\prime p}+d_{4}\brpsip_{B}\brgamma_{A}\psip_{C}\\
{}&+\ddp_{1}\brrho\gamma_{ABC}\rho+\ddp_{2}\brrho\gamma_{BC}\psi_{A}+\ddp_{3}\brpsi_{\brp}\gamma_{ABC}\psi^{\brp}+\ddp_{4}\brpsi_{B}\gamma_{A}\psi_{C}\,,
\ea
\label{letcoeff}
\ee
of which the coefficients  must satisfy the following nine constraints,
\be
\ba{ll}
\aap_{4}+\bbp_{4}=4\ccp_{1}\,,~~~~&~~~~
\aap_{2}+\ccp_{2}=\aap_{2}-\half\aap_{4}\,,\\
\aap_{4}+\ccp_{4}=-4\ccp_{3}\,,~~~~&~~~~
\aap_{4}+\ddp_{4}=-2(\aap_{2}-\half\aap_{4})\,,\\
a_{1}+b_{3}=\textstyle{\frac{1}{16}}(a_{2}-d_{2})\,,~~~~&~~~~
a_{3}+c_{1}=-\textstyle{\frac{1}{16}}(a_{2}-d_{2})\,,\\
a_{3}-c_{3}=\textstyle{\frac{1}{6}}(c_{4}-a_{4})\,,~~~~&~~~~
a_{1}+d_{1}=-\textstyle{\frac{1}{2}}(a_{2}+b_{2})\,,\\
a_{3}+d_{3}=\textstyle{\frac{1}{2}}(a_{2}+c_{2})\,.~~~~&~~~~
\ea
\label{9const}
\ee
A particularly simple solution is given by 
\be
\ba{ll}
\bbp_{1}=\ccp_{1}=\ddp_{1}=-\half(\aap_{2}-\half\aap_{4})\,,~~~&~~
\bbp_{2}=\ccp_{2}=\ddp_{2}=-\half\aap_{4}\,,\\
\bbp_{3}=\ccp_{3}=\ddp_{3}=\half(\aap_{2}-\half\aap_{4})\,,~~~&~~~
\bbp_{4}=\ccp_{4}=\ddp_{4}=-2\aap_{2}\,,
\ea
\label{bcdsol}
\ee
and
\be
\ba{ll}
b_{1}=c_{1}=d_{1}=-\textstyle{\frac{1}{9}}(a_{1}+a_{2}+8a_{3})\,,~~~&~~
b_{2}=c_{2}=d_{2}=-\textstyle{\frac{1}{9}}(16a_{1}+7a_{2}-16a_{3})\,,\\
b_{3}=c_{3}=d_{3}=-\textstyle{\frac{1}{9}}(8a_{1}-a_{2}+a_{3})\,,~~~&~~
b_{4}=c_{4}=d_{4}=\textstyle{\frac{1}{3}}(16a_{1}-2a_{2}+20a_{3}+3a_{4})\,.
\ea
\label{bcdsol}
\ee
Specifically, for   $\Gammas_{ABC}$ and $\Gammaps_{ABC}$ given in (\ref{GsGps}) as
\be
\ba{llllllll}
a_{1}=-i\textstyle{\frac{11}{96}}\,,~&
a_{2}=i\textstyle{\frac{5}{4}}\,,~&
a_{3}=i\textstyle{\frac{5}{24}}\,,~&
a_{4}=-2i\,,~&
\aap_{1}=0\,,~&
\aap_{2}=i\textstyle{\frac{5}{2}}\,,~&
\aap_{3}=0\,,~&
\aap_{4}=0\,,
\ea
\ee
we achieve  (\ref{GGG}),
\be
\ba{ll}
\Gammash_{ABC}=\Gammafl_{ABC}=\tGamma_{ABC}=\!&\Gamma_{ABC}
-i\textstyle{\frac{23}{54}}\brrho\gamma_{ABC}\rho\,
+i\textstyle{\frac{23}{27}}\brrho\gamma_{BC}\psi_{A}\,
+i\textstyle{\frac{23}{54}}\brpsi^{\brp}\gamma_{ABC}\psi_{\brp}\,
-i\textstyle{\frac{73}{18}}\brpsi_{B}\gamma_{A}\psi_{C}\\
{}&\quad-i\textstyle{\frac{5}{4}}\brrhop\brgamma_{ABC}\rhop
+i\textstyle{\frac{5}{2}}\brrhop\brgamma_{BC}\psip_{A}
+i\textstyle{\frac{5}{4}}\brpsi^{\prime p}\brgamma_{ABC}\psip_{p}
-5i\brpsip_{B}\brgamma_{A}\psip_{C}\,,\\
\Gammapsh_{ABC}=\Gammapfl_{ABC}=\tGammap_{ABC}=\!&\Gamma_{ABC}
-i\textstyle{\frac{23}{54}}\brrhop\brgamma_{ABC}\rhop
+i\textstyle{\frac{23}{27}}\brrhop\brgamma_{BC}\psip_{A}
+i\textstyle{\frac{23}{54}}\brpsi^{\prime p}\brgamma_{ABC}\psip_{p}
-i\textstyle{\frac{73}{18}}\brpsip_{B}\brgamma_{A}\psip_{C}\\
{}&\quad-i\textstyle{\frac{5}{4}}\brrho\gamma_{ABC}\rho\,
+i\textstyle{\frac{5}{2}}\brrho\gamma_{BC}\psi_{A}\,
+i\textstyle{\frac{5}{4}}\brpsi^{\brp}\gamma_{ABC}\psi_{\brp}\,
-5i\brpsi_{B}\gamma_{A}\psi_{C}\,.
\ea
\label{GGG2}
\ee
Alternatively, in a similar fashion to  Ref.\cite{Jeon:2011sq}, we might  set
\be
\ba{l}
\tGamma_{ABC}=\Gamma_{ABC}
-i\textstyle{\frac{17}{48}}\brrho\gamma_{ABC}\rho+i\textstyle{\frac{5}{2}}\brrho\gamma_{BC}\psi_{A}+i\textstyle{\frac{1}{4}}\brpsi^{\brp}\gamma_{ABC}\psi_{\brp} -5i \brpsi^{\prime}_{B} \brgamma_{A} \psi^{\prime}_{C}\,,\\
\tGamma^{\prime}_{ABC}=\Gamma_{ABC}
-i\textstyle{\frac{17}{48}}\brrhop\brgamma_{ABC}\rhop+i\textstyle{\frac{5}{2}}\brrhop\brgamma_{BC}\psip_{A}+i\textstyle{\frac{1}{4}}\brpsi^{\prime p}\brgamma_{ABC}\psip_{p}-5i \brpsi_{B}\gamma_{A} \psi_{C}\,,\\
\Gammash_{ABC}=\Gamma_{ABC}
+i\textstyle{\frac{17}{24}}\brrho\gamma_{BC}\psi_{A}+i\textstyle{\frac{31}{96}}\brpsi^{\brr}\gamma_{ABC}\psi_{\brr}\,,\\
\Gammapsh_{ABC}=\Gamma_{ABC}
+i\textstyle{\frac{17}{24}}\brrhop\brgamma_{BC}\psi^{\prime}_{A}+i\textstyle{\frac{31}{96}}\brpsi^{\prime r}\brgamma_{ABC}\psi^{\prime}_{r}\,,\\
\Gamma^{\flat}_{ABC} =\Gamma_{ABC}
-i\textstyle{\frac{31}{96}}\brrho\gamma_{ABC}\rho+i\textstyle{\frac{1}{2}}\brrho\gamma_{BC}\psi_{A}+i\textstyle{\frac{5}{12}}\brpsi^{\brp}\gamma_{ABC}\psi_{\brp} -4i \brpsi_{B}\gamma_{A} \psi_{C} +i \frac{5}{2} \brrho^{\prime} \brgamma_{BC} \psi^{\prime}_{A} \,,
\\
\Gamma^{\prime\flat}_{ABC} =\Gamma_{ABC}
-i\textstyle{\frac{31}{96}}\brrhop\brgamma_{ABC}\rhop+i\textstyle{\frac{1}{2}}\brrhop\brgamma_{BC}\psip_{A}
+i\textstyle{\frac{5}{12}}\brpsi^{\prime p}\brgamma_{ABC}\psip_{p} -4i \brpsi^{\prime}_{B} \brgamma_{A} \psi^{\prime}_{C}+i\frac{5}{2} \brrho \gamma_{BC} \psi_{A}\,.
\ea
\ee
That is to say, there are various ways of absorbing   the higher order fermionic terms  into the torsions as long as the constraint (\ref{9const}) is satisfied. In this paper, we choose (\ref{GGG2}) and hence (\ref{GGG}) such that,  the entire  equations of motion~(\ref{varL}) can be written in terms of only   two kind of torsions: one for the unprimed fermions and the other for the primed fermions.

\newpage

\section{$\cN=2$ supersymmetric invariance of the action\label{secAppsusy}}
Here, we sketch  our verification of   the $\cN=2$ supersymmetric invariance of the action as in (\ref{varLSUSY}), order by order in fermions.  We substitute the  ${\cN=2}$ supersymmetry transformation rules (\ref{N2SUSY})  into ~(\ref{varL})  and organize the   supersymmetric variation  of the Lagrangian as
\be
\deltaS\cL_{\typeT}=\deltaS\cL^{[1]}_{\typeT}+\deltaS\cL^{[3]}_{\typeT}+\deltaS\cL^{[5]}_{\typeT}\,,
\ee
where $\deltaS\cL^{[1]}_{\typeT}$, $\deltaS\cL^{[3]}_{\typeT}$ and $\deltaS\cL^{[5]}_{\typeT}$ denote  respectively the linear, cubic and quintic order  terms in fermions which are DFT-dilatinos and  gravitinos. \\

First of all, we focus on the linear order terms which     decompose  into four parts:
\be
\deltaS\cL^{[1]}_{\typeT}\seceq\Delta_{\rho}+\Delta_{\psi}+\Delta_{\cF}+
\Delta_{\cF^{2}}\,,
\ee
where, disregarding the total derivative terms, we have
\be
\ba{ll}
\Delta_{\rho}=-ie^{-2d}\Big[&\!\!\textstyle{\frac{1}{8}}(\brrho\varepsilon+\brrhop\varepsilonp)
(P^{AB}P^{CD}-\brP^{AB}\brP^{CD})\So_{ACBD}\\
{}&
+\brrho\left(\half\gamma^{pq}[\cDo_{p},\cDo_{q}]+\cDo{}^{A}\cDo_{A}\right)\varepsilon
-\brrhop\left(\half\brgamma^{\brp\brq}[\cDo_{\brp},\cDo_{\brq}]+\cDo{}^{A}\cDo_{A}\right)\varepsilonp\Big]\,,\\
\multicolumn{2}{l}{\Delta_{\psi}=-ie^{-2d}\Big[\,~\half(\brvarepsilon\gamma^{p}\psi^{\brq}
-\brvarepsilonp\brgamma^{\brq}\psi^{\prime p})\So_{p\brq}+\brpsi^{\brq}\gamma^{p}[\cDo_{p},\cDo_{\brq}]\varepsilon+
\brpsi^{\prime p}\brgamma^{\brq}[\cDo_{p},\cDo_{\brq}]\varepsilonp\Big]\,,}\\
\Delta_{\cF}=-ie^{-2d}\Big[&\brrho(\cDo_{\brp}\cF)\brgamma^{\brp}\varepsilonp-\brrhop(\cDo_{p}\brcF)\gamma^{p}\varepsilon+
\brpsi_{\brp}(\gamma^{p}\cDo_{p}\cF)\brgamma^{\brp}\varepsilonp-
\brpsi^{\prime}_{p}(\brgamma^{\brp}\cDo_{\brp}\brcF)\gamma^{p}\varepsilon\\
{}&+\half\Tr\!\left[(\rho\brvarepsilonp-\varepsilon\brrhop+
\gamma^{p}\varepsilon\brpsi^{\prime}_{p}-\brpsi_{\brp}\brvarepsilonp\brgamma^{\brp})\overline{\cDo_{-}\cF}\,\right]\Big]\,,\\
\multicolumn{2}{l}{\Delta_{\cF^{2}}=+ie^{-2d}\Big[~
\brpsi_{\brp}\gamma_{q}\cF\brgamma^{\brp}\brcF\gamma^{q}\varepsilon
-\brpsi^{\prime}_{p}\brgamma_{\brq}\brcF\gamma^{p}\cF\brgamma^{\brq}\varepsilonp
+\half(\brvarepsilon\gamma_{p}\psi_{\brq}-\brvarepsilonp\brgamma_{\brq}\psi^{\prime}_{p})\Tr\!\left(\gamma^{\eleven}\gamma^{p}\cF\brgamma^{\brq}\brcF\right)
\Big]\,.}
\ea
\ee

We show, up to the level matching section constraint~(\ref{constraint}), each of them vanishes except the last one, $\Delta_{\cF^{2}}$.
\begin{enumerate}
\item For $\Delta_{\rho}$.\\
We first  note 
\be
\ba{l}
{}[\cDo_{A},\cDo_{B}]\varepsilon=F_{AB}\varepsilon-\Gamma^{C}{}_{AB}\cDo_{C}\varepsilon\,,\\
{}[\cDo_{A},\cDo_{B}]\varepsilonp=\brF_{AB}\varepsilonp-\Gamma^{C}{}_{AB}\cDo_{C}\varepsilonp\,,\\
\cDo_{A}\cDo{}^{A}\varepsilon=(\partial_{A}\Phio{}^{A}+\Gamma_{A}{}^{AB}\Phio_{B}-
\Phio_{A}\Phio{}^{A})\varepsilon+2\Phio_{A}\cDo{}^{A}\varepsilon\,,\\
\cDo_{A}\cDo{}^{A}\varepsilonp=(\partial_{A}\brPhio{}^{A}+\Gamma_{A}{}^{AB}\brPhio_{B}-
\brPhio_{A}\brPhio{}^{A})\varepsilonp+2\brPhio_{A}\cDo{}^{A}\varepsilonp\,.
\ea
\ee
Then, due to the identities~\cite{Jeon:2012kd},
\be
\ba{l}
\partial_{A}\Phio{}^{A}+\Phio_{A}\Phio{}^{A}+\half\gamma^{AB}F_{AB}+\left(\Gamma_{B}{}^{BA}-\half\Gamma^{A}{}_{pq}\gamma^{pq}\right)\Phio_{A}\seceq-\quarter \So_{ABCD}P^{AC}P^{BD}\,,\\
\partial_{A}\brPhio{}^{A}+\brPhio_{A}\brPhio{}^{A}+\half\brgamma^{AB}\brF_{AB}+\left(\Gamma_{B}{}^{BA}-\half\Gamma^{A}{}_{\brp\brq}\brgamma^{\brp\brq}\right)\brPhio_{A}\seceq-\quarter \So_{ABCD}\brP^{AC}\brP^{BD}\,,
\ea
\ee
we obtain (\textit{c.f.~}\cite{Coimbra:2011nw,Hohm:2011nu})
\be
\ba{l}
{}\left(\half\gamma^{pq}[\cDo_{p},\cDo_{q}]+\cDo{}^{A}\cDo_{A}\right)\varepsilon
\seceq-\quarter P^{AB}P^{CD}\So_{ACBD}\varepsilon\,,\\
\left(\half\brgamma^{\brp\brq}[\cDo_{\brp},\cDo_{\brq}]+\cDo{}^{A}\cDo_{A}\right)\varepsilonp\seceq-\quarter \brP^{AB}\brP^{CD}\So_{ACBD}\varepsilonp\,.
\ea
\ee
These simplify  $\Delta_{\rho}$ as
\be
\Delta_{\rho}\seceq i\textstyle{\frac{1}{8}}e^{-2d}
(\brrho\varepsilon-\brrhop\varepsilonp)
(P^{AB}P^{CD}+\brP^{AB}\brP^{CD})\So_{ACBD}\,,
\ee
and finally from the identity~\cite{Jeon:2011cn,Jeon:2012kd},
\be
(P^{AB}P^{CD}+\brP^{AB}\brP^{CD})\So_{ACBD}\seceq 0\,,
\ee
we  note $\Delta_{\rho}\seceq0$.

\item For $\Delta_{\psi}$.\\
From
\be
\ba{ll}
{}[\cDo_{p},\cDo_{\brq}]\varepsilon\seceq\half \So_{p\brq rs}\gamma^{rs}\varepsilon\,,~~~~&~~~~
{}[\cDo_{p},\cDo_{\brq}]\varepsilonp\seceq\half \So_{p\brq\brr\brs}\brgamma^{\brr\brs}\varepsilonp\,,
\ea
\ee
$\Delta_{\psi}$ reduces to
\be
\Delta_{\psi}\seceq -i\half e^{-2d}(\brpsi^{\prime p}\brgamma^{\brq}\varepsilonp+
\brpsi^{\brq}\gamma^{p}\varepsilon)(P^{AB}-\brP^{AB})\So_{pA\brq B}\,.
\ee
Then, from the identity~\cite{Jeon:2011cn,Jeon:2012kd}, 
\be
(P^{AB}-\brP^{AB})\So_{pA\brq B}\seceq 0\,,
\ee 
we verify $\Delta_{\psi}\seceq 0$.
\item For $\Delta_{\cF}$.\\
Straightforward computation may give
\be
\ba{ll}
\Delta_{\cF}=-ie^{-2d}\Big[&\!\!\!\!
\brrho(1-\gamma^{\eleven})\cDo_{\brp}\cF\brgamma^{\brp}\varepsilonp+
\brvarepsilon(1+\gamma^{\eleven})\gamma^{p}\cDo_{\brq}\cF\brgamma^{\brq}\psi_{p}^{\prime}\\
{}&-\half\Tr\!\left[
(\rhop\brvarepsilon+\psi^{\prime}_{p}\brvarepsilon\gamma^{p}+
\varepsilonp\brrho+\brgamma^{\brp}\varepsilonp\psi_{\brp})
\cDo_{+}\cF\right]\Big]\,.
\ea
\ee

Hence, from the chirality of the fermions and  the nilpotent property~\cite{Jeon:2012kd}, 
\be
\cDo_{+}\cF=(\cDo_{+})^{2}\cC\seceq 0\,,
\ee
we  note $\Delta_{\cF}\seceq 0$.
\item For $\Delta_{\cF^{2}}$.\\
Using  the well-known Fierz identities involving  a cyclic sum over three spinorial 
indices (\ref{FierzYM}), it is easy to  show that  $\Delta_{\cF^{2}}$ (and hence $\deltaS\cL^{[1]}_{\typeT}$) reduces  to
\be
\deltaS\cL^{[1]}_{\typeT}\seceq\Delta_{\cF^{2}}=i\quarter e^{-2d}(\brvarepsilon\gamma_{p}\psi_{\brq}
-\brvarepsilonp\brgamma_{\brq}\psi^{\prime}_{p})\Tr[\gamma^{p}(1-\gamma^{\eleven})\cF\brgamma^{\brq}\brcF]\,.
\label{L1}
\ee
\end{enumerate}
~\\

We now turn to the higher order terms in fermions.  After long and tedious computations, using the various  Fierz identities presented  in Appendix~\ref{secAppFI},  we obtain, for the cubic order terms, 
\be
\deltaS\cL^{[3]}_{\typeT}\seceq\half  e^{-2d}(\brvarepsilon\gamma_{p}\psi_{\brq}
-\brvarepsilonp\brgamma_{\brq}\psi^{\prime}_{p})\left(\brrho\gamma^{p}\cF\brgamma^{\brq}\rhop-
\brpsi_{\brs}\gamma^{t}\gamma^{p}\cF\brgamma^{\brq}\brgamma^{\brs}\psi^{\prime}_{t}\right)\,,
\label{L3}
\ee
and  for the quintic order terms,
\be
\deltaS\cL^{[5]}_{\typeT}=i\quarter e^{-2d}(\brvarepsilon\gamma_{p}\psi_{\brq}
-\brvarepsilonp\brgamma_{\brq}\psi^{\prime}_{p})\left(\brrho\gamma^{p}\gamma^{s}\brpsi_{\brs}\,\brpsi^{\prime}_{s}\brgamma^{\brs}\brgamma^{\brq}\rho^{\prime}+
\half\brpsi_{\brs}\gamma^{s}\gamma^{p}\gamma^{t}\psi_{\brt}\,
\brpsi^{\prime}_{s}\brgamma^{\brs}\brgamma^{\brq}\brgamma^{\brt}\psi^{\prime}_{t}\right)\,.
\label{L5}
\ee
In fact, the cubic order terms decompose into two parts: one involving the R-R field strength, $\cF$, and the other with the torsionless master derivative, $\cDo_{A}$. The former reduces to (\ref{L3}) and the latter turns out to be a  total derivative which we neglect. The computation of the quintic order terms is genuinely algebraic. \\

\textit{At last,}  adding up (\ref{L1}), (\ref{L3}) and (\ref{L5}), we obtain the final expression~(\ref{varLSUSY}). This completes our   verification of the $\cN=2$ supersymmetric invariance of the  action,  modulo the self-duality~(\ref{SD}), to the full order in fermions.\\
\newpage

\section{Fierz identities\label{secAppFI}}
With the chiral/anti-chiral  projections, 
\be
\ba{ll}
\gamma_{\pm}:=\half\left(1\pm\gamma^{\eleven}\right)\,,~~~~&~~~~
\brgamma_{\pm}:=\half\left(1\pm\brgamma^{\eleven}\right)\,,
\ea
\ee
where
\be
\ba{ll}
\gamma^{\eleven}=\gamma^{012\cdots 9}\,,~~~~&~~~~
\brgamma^{\eleven}=\brgamma^{012\cdots 9}\,,
\ea
\ee
relevant  Fierz identities  are   as follows (\textit{c.f.~}\cite{Jeon:2011sq}). 
\begin{equation}
\ba{l}
(\gamma_{-})^{\alpha}_{\phantom{11}\lambda}(\gamma_{+})^{\delta}_{\phantom{11}\beta}=
{\textstyle\frac{1}{32\times 5!}}(\gamma_{a_{5}\cdots a_{1}} \gamma_{-})^{\delta}_{\phantom{11}\lambda}(\gamma^{a_{1}\cdots a_{5}} \gamma_{+})^{\alpha}_{\phantom{11}\beta}+
\dis{\sum_{n=1, 3}}{\textstyle\frac{1}{16\times n!}}(\gamma_{a_{n}\cdots a_{1}} \gamma_{-})^{\delta}_{\phantom{11}\lambda}(\gamma^{a_{1}\cdots a_{n}} \gamma_{+})^{\alpha}_{\phantom{11}\beta}\, ,\\
(\gamma_{\pm})^{\alpha}_{\phantom{11}\lambda}(\gamma_{\pm})^{\delta}_{\phantom{11}\beta}=\dis{\sum_{n=0, 2, 4}}{\textstyle\frac{1}{16\times n!}}(\gamma_{a_{n}\cdots a_{1}} \gamma_{\pm})^{\delta}_{\phantom{11}\lambda}(\gamma^{a_{1}\cdots a_{n}} \gamma_{\pm})^{\alpha}_{\phantom{11}\beta}\,,
\ea
\end{equation}
\be
\gamma_{a_{m}\cdots a_{1}}\gamma^{b_{1}\cdots b_{n}}=
\sum_{l=0}^{{\rm{min}}[m,n]}l!\binom{m}{l}\binom{n}{l}\gamma_{[a_{m}\cdots a_{l+1}}{}^{[b_{l+1}\cdots b_{n}}\delta_{a_{1}}^{~b_{1}}\cdots\delta_{a_{l}]}^{~b_{l}]}\,,
\ee

\be
\ba{l}
\gamma_{p}\gamma_{a_{1}\cdots a_{n}}\gamma^{p}=(-1)^n (10-2n)\gamma_{a_{1}\cdots a_{n}}\,,\\
\gamma_{pq}\gamma_{a_{1}\cdots a_{n}}\gamma^{pq}=(-90+40n-4n^2)\gamma_{a_{1}\cdots a_{n}}\,,\\
\gamma_{pqr}\gamma_{a_{1}\cdots a_{n}}\gamma^{pqr}=(-1)^{n}(-720+544n-120n^{2}+8n^{3})\gamma_{a_{1}\cdots a_{n}}\,,
\ea
\ee
\be
\ba{ll}
\left(C_{+}\gamma^{p}\gamma_{\pm}\right)_{(\alpha\beta}
\left(C_{+}\gamma_{p}\gamma_{\pm}\right)_{\gamma)\delta}=0\,,~~~~&~~~~
\left(\brC_{+}\brgamma^{\brp}\brgamma_{\pm}\right)_{(\bralpha\brbeta}
\left(\brC_{+}\brgamma_{\brp}\brgamma_{\pm}\right)_{\brgamma)\brdelta}=0\,.
\ea
\label{FierzYM}
\ee
In particular, 
\be
\gamma_{abc}\gamma^{stu}=-\gamma^{stu}\gamma_{abc}+18\gamma^{[st}\gamma_{[ab}\delta^{u]}_{c]}+72\gamma^{[s}\gamma_{[a}\delta^{tu]}_{bc]}-48\delta^{~s}_{u}\delta^{~t}_{b}\delta^{~u}_{c}\,,
\ee
\be
\gamma^{pq}\gamma_{abc}\gamma^{stu}\gamma_{pq}=-6\gamma^{stu}\gamma_{abc}+108\gamma^{[st}\gamma_{[ab}\delta^{u]}_{c]}-144\gamma^{[s}\gamma_{[a}\delta^{tu]}_{bc]}-864\delta^{~s}_{u}\delta^{~t}_{b}\delta^{~u}_{c}\,.
\ee
~\\

For the  chiral gravitinos,  $\psi_{\brp}=+\gamma^{\eleven}\psi_{\brp}$, and the anti-chiral DFT-dilatinos, $\rho=-\gamma^{\eleven}\rho$,
\be
\ba{ll}
\bar{\rho}\gamma^{pqr}\rho(\bar{\rho}\gamma_{pq})_{\alpha}=0\,,~~~~&~~~~\bar{\rho}\gamma^{pqr}\rho(\bar{\rho}\gamma_{pqr})_{\alpha}=0\,,
\ea
\label{r4}
\ee

\be
\ba{ll}
\rho\brrho=\textstyle{\frac{1}{96}}(\brrho\gamma^{pqr}\rho)\gamma_{-}\gamma_{pqr}\gamma_{+}\,,~~~&~~~
\psi^{\brp}\brpsi_{\brp}=\textstyle{\frac{1}{96}}(\brpsi^{\brp}\gamma^{pqr}\psi_{\brp})\gamma_{+}\gamma_{pqr}\gamma_{-}\,,
\ea
\ee

\be
\textstyle{\frac{1}{16}}\brrho\gamma^{pqr}\rho\,\bar{\psi}_{\brp}\gamma_{pqr}\psi^{\brp}=
-\bar{\rho}\gamma_{pq}\psi_{\brp}\,\brrho\gamma^{pq}\psi^{\brp}\,,
\label{rrpp}
\ee

\be
\ba{ll}
\gamma^{pqrst}\varepsilon\brpsi_{\brp}\gamma_{pqrst}=0\,,~~~&~~~
\gamma^{pqrst}\rho\brpsi_{\brp}\gamma_{a}\gamma_{pqrst}=0\,,
\ea
\ee

\be
 \half \gamma^{q} \psi_{\brp} \brvarepsilon \gamma_{pq} \rho-\half \gamma^{q} \varepsilon\, \brpsi_{\brp} \gamma_{pq} \rho +\textstyle{ \frac{1}{8}} \gamma_{p} \gamma^{qr} \varepsilon \,\brpsi_{\brp} \gamma_{qr}\rho + \half \gamma_{p}\psi_{\brp}\brvarepsilon \rho +\quarter \gamma_{p}\varepsilon\brpsi_{\brp} \rho + \rho\brvarepsilon \gamma_{p} \psi_{\brp}  =0 \,,
\label{fi6terms}
\ee

\be
\brpsi_{\brs}\gamma^{s}\gamma^{p}\gamma^{t}\psi_{\brt}\,
\brpsi^{\prime}_{s}\brgamma^{\brs}\brgamma^{\brq}\brgamma^{\brt}\psi^{\prime}_{t}=
\brpsi_{\brs}\gamma^{pst}\psi^{\brs}\brpsi^{\prime}_{s}\brgamma^{\brq}\psi^{\prime}_{t}
-2\brpsi^{\brq}\gamma^{pst}\psi_{\brs}\brpsi^{\prime}_{s}\brgamma^{\brs}\psi^{\prime}_{t}
+\brpsi^{\prime}_{s}\brgamma^{\brq\brs\brt}\psi^{\prime s}\brpsi_{\brs}\gamma^{p}\psi_{\brt}
-2\brpsi^{\prime p}\brgamma^{\brq\brs\brt}\psi^{\prime}_{s}\brpsi_{\brs}\gamma^{s}\psi_{\brt}\,.
\label{fi5terms}
\ee
Further, with an additional anti-chiral fermion, $\chi=-\gamma^{\eleven}\chi$, 
\be
\textstyle{\frac{1}{96}}\brvarepsilon\gamma^{rst}\psi_{\brp}\brrho\gamma_{rst}\chi
-\textstyle{\frac{3}{2}}\brrho\varepsilon\brpsi_{\brp}\chi
-\textstyle{\frac{1}{8}}\brvarepsilon\gamma^{rs}\chi\brrho\gamma_{rs}\psi_{\brp}
-\textstyle{\frac{1}{4}}\brrho\psi_{\brp}\brvarepsilon\chi
+\half\brvarepsilon\gamma^{r}\psi_{\brp}\brrho\gamma_{r}\chi=0\,.
\label{fichi}
\ee
With the R-R field strength, $\cF=\mp\gamma^{\eleven}\cF\brgamma^{\eleven}$,
\begin{equation}
\ba{l}
(\gamma_{p}\psi_{\bar{q}})(\bar{\varepsilon}\gamma^{p}{\cal F}\bar{\gamma}^{\bar{q}})=-\half (\bar{\psi}_{\brq}\gamma_{p}\varepsilon)(\gamma^{p}{\cal F}\bar{\gamma}^{\bar{q}})-{\textstyle\frac{1}{24}}(\bar{\psi}_{\bar{q}}\gamma_{abc}\varepsilon)(\gamma^{abc}{\cal F}\bar{\gamma}^{\bar{q}})\,,\\
(\bar{\varepsilon}\gamma_{p}\psi_{\bar{q}})(\gamma^{p}{\cal F}\bar{\gamma}^{\bar{q}})=-\half(\gamma^{p}\varepsilon)(\bar{\psi}_{\bar{q}}\gamma_{p}{\cal F}\bar{\gamma}^{\bar{q}})+{\textstyle\frac{1}{24}}(\gamma^{abc}\varepsilon)(\bar{\psi}_{\bar{q}}\gamma_{abc}{\cal F}\bar{\gamma}^{\bar{q}})\,,\\
(\bar{\varepsilon}\gamma_{abc}\psi_{\bar{q}})(\gamma^{abc}{\cal F}\bar{\gamma}^{\bar{q}})=-18(\gamma^{p}\varepsilon)(\bar{\psi}_{\bar{q}}\gamma_{p}{\cal F}\bar{\gamma}^{\bar{q}})-\half(\gamma^{abc}\varepsilon)(\bar{\psi}_{\bar{q}}\gamma_{abc}{\cal F}\bar{\gamma}^{\bar{q}})\,,\\
(\gamma^{p}{\cal F}\bar{\gamma}^{\bar{q}}\varepsilon^{\prime})(\bar{\psi}^{\prime}_{p}\bar{\gamma}_{\bar{q}})=\half (\gamma^{p}{\cal F}\bar{\gamma}^{\bar{q}})(\bar{\psi}^{\prime}_{p}\bar{\gamma}_{\bar{q}}\varepsilon^{\prime})-{\textstyle\frac{1}{24}}(\gamma^{p}{\cal F}\bar{\gamma}^{\bra\bar{b}\bar{c}})(\bar{\psi}^{\prime}_{p}\bar{\gamma}_{\bar{a}\bar{b}\bar{c}}\varepsilon^{\prime})\,,\\
(\gamma^{p}{\cal F}\bar{\gamma}^{\bar{q}})(\bar{\psi}^{\prime}_{p}\bar{\gamma}_{\bar{q}}\varepsilon^{\prime})=-\half(\gamma^{p}{\cal F}\bar{\gamma}_{\bar{q}}\psi^{\prime}_{p})(\bar{\varepsilon}^{\prime}\bar{\gamma}^{\bar{q}})+{\textstyle\frac{1}{24}}(\gamma^{p}{\cal F}\bar{\gamma}_{\bar{a}\bar{b}\bar{c}}\psi^{\prime}_{p})(\bar{\varepsilon}^{\prime}\bar{\gamma}^{\bra\bar{b}\bar{c}})\,,\\
(\gamma^{p}{\cal F}\bar{\gamma}^{\bra\bar{b}\bar{c}})(\bar{\psi}^{\prime}_{p}\bar{\gamma}_{\bar{a}\bar{b}\bar{c}}\varepsilon^{\prime})=-18(\gamma^{q}{\cal F}\bar{\gamma}_{\bar{p}}\psi^{\prime}_{q})(\bar{\varepsilon}^{\prime}\bar{\gamma}^{\bar{p}})-{\textstyle\frac{1}{2}}(\gamma^{p}{\cal F}\brgamma_{\bar{a}\bar{b}\bar{c}}\psi_{p}^{\prime})(\bar{\varepsilon}^{\prime}\bar{\gamma}^{\bra\bar{b}\bar{c}})\,.
\ea
\end{equation}

\end{document}